\pdfoutput=1
\newcommand*{\ATLASLATEXPATH}{latex/}
\documentclass[cernpreprint,txfonts,UKenglish,texlive=2016]{\ATLASLATEXPATH atlasdoc}
\usepackage[biblatex=true]{\ATLASLATEXPATH atlaspackage}
\usepackage{\ATLASLATEXPATH atlasphysics}
\usepackage{mathrsfs}
\usepackage{multirow}
\usepackage{rotating}
\usepackage{longtable}
\usepackage{tabularx}
\usepackage{ltablex}
\usepackage{graphicx}
\usepackage{caption}
\usepackage{booktabs,siunitx}
\usepackage{\ATLASLATEXPATH atlasbiblatex}
\usepackage{\ATLASLATEXPATH atlascontribute}
\graphicspath{{logos/}{figures/}}
\AtlasTitle{Measurement of the $W$ boson polarisation in $t\bar{t}$ events from $pp$ collisions at $\sqrt{s}$ = 8 TeV in the lepton~+~jets channel with ATLAS}
\author{The ATLAS Collaboration}
\AtlasRefCode{TOPQ-2016-02}
\PreprintIdNumber{CERN-PH-2016-219}
\AtlasJournalRef{Eur. Phys. J. C 77 (2017) 264}
\AtlasDOI{10.1140/epjc/s10052-017-4819-4}

\AtlasAbstract{%
  This paper presents a measurement of the polarisation of $W$
  bosons from $t\bar{t}$ decays, reconstructed in events with one high-$p_{\mbox{T}}$ lepton and at least four jets. Data from $pp$ collisions at the LHC were 
  collected at $\sqrt{s}$ = 8 TeV and correspond to an integrated luminosity 
  of 20.2~\mbox{fb$^{-1}$}.   The angle $\theta^{*}$ between the $b$-quark from the top quark decay and a direct $W$ boson decay product in the $W$ boson rest frame is sensitive to the $W$ boson polarisation. Two different $W$ decay products are used as polarisation analysers: the 
  charged lepton and the down-type quark for the leptonically and 
  hadronically decaying $W$ boson, respectively. The most precise measurement of the $W$ boson polarisation via the distribution of $\cos{\theta^{*}}$ is obtained using the leptonic analyser and events in which at least two of the jets are tagged as $b$-quark jets.
  The fitted fractions of longitudinal, left- and right-handed polarisation states are $F_{\mathrm{0}}=~0.709~\pm~{0.019}$, $F_{\mathrm{L}}=~0.299~\pm~{0.015}$ and $F_{\mathrm{R}}=~-0.008~\pm~{0.014}$, and are the most precisely measured $W$ boson polarisation fractions to date. Limits on anomalous couplings of the $Wtb$ vertex are set.}

\hypersetup{pdftitle={ATLAS document},pdfauthor={The ATLAS Collaboration}}
\newcommand{\AtlasCoordFootnote}{%
ATLAS uses a right-handed coordinate system with its origin at the nominal interaction point (IP)
in the centre of the detector and the $z$-axis along the beam pipe.
The $x$-axis points from the IP to the centre of the LHC ring,
and the $y$-axis points upwards.
Cylindrical coordinates $(r,\phi)$ are used in the transverse plane, 
$\phi$ being the azimuthal angle around the $z$-axis.
The pseudorapidity is defined in terms of the polar angle $\theta$ as $\eta = -\ln \tan(\theta/2)$.
The angular distance is measured in units of $\Delta R \equiv \sqrt{(\Delta\eta)^{2} + (\Delta\phi)^{2}}$.}
\newcommand{\mtw}{m_{\text{T}}(W)}
\newcommand*{\costheta}{\ensuremath{\cos{\theta^{*}}}\xspace}

\newcommand*{\FN}{\ensuremath{F_{\mathrm{0}}}\xspace}
\newcommand*{\FL}{\ensuremath{F_{\mathrm{L}}}\xspace}
\newcommand*{\FR}{\ensuremath{F_{\mathrm{R}}}\xspace}
\newcommand*{\Fi}{\ensuremath{F_{i}}\xspace}
\newcommand*{\Wbos}{\Wboson~boson\xspace}
\newcommand*{\dQ}{down-type quark\xspace}

\newcommand*{\com}{centre-of-mass\xspace}
\newcommand*{\ljets}{lepton+jets\xspace}
\newcommand*{\ejets}{\ensuremath{e} + jets\xspace}
\newcommand*{\mujets}{\ensuremath{\mu} + jets\xspace}
\newcommand*{\powpy}{\textsc{Powheg+Pythia}\xspace}
\newcommand*{\powhw}{\textsc{Powheg+Herwig}\xspace}
\newcommand*{\mtop}{\ensuremath{m_{t}}\xspace}

\newcommand*{\dR}{\ensuremath{\Delta R \xspace}}
\newcommand*{\etaclus}{\ensuremath{\eta_{\text{cluster}}}\xspace}
\newcommand*{\absetaclus}{\ensuremath{|\eta_{\text{cluster}}|}\xspace}
\newcommand*{\Zee}{\ensuremath{Z\rightarrow ee}\xspace}
\newcommand*{\bt}{\ensuremath{b}\xspace}
\newcommand*{\wjets}{\ensuremath{W}+jets\xspace}
\newcommand*{\focent}{\ensuremath{0.709}\xspace}%
\newcommand*{\fostat}{\ensuremath{0.012}\xspace}%
\newcommand*{\fosystplus}{\ensuremath{0.015}\xspace}%
\newcommand*{\flcent}{\ensuremath{0.299}\xspace}%
\newcommand*{\flstat}{\ensuremath{0.008}\xspace}%
\newcommand*{\flsystplus}{\ensuremath{0.013}\xspace}%
\newcommand*{\frcent}{\ensuremath{-0.008}\xspace}%
\newcommand*{\frstat}{\ensuremath{0.006}\xspace}%
\newcommand*{\frsystplus}{\ensuremath{0.012}\xspace}%
\newcommand*{\statBG}{\mbox{$\;$(stat.+bkg. norm.)}\xspace}

\begin{document}
\maketitle
\clearpage
\section{Introduction}
\label{sec:intro}

The top quark, discovered in 1995 by the CDF and D0 collaborations~\cite{top_disc_CDF, top_disc_D0} is the heaviest known elementary particle. 
It decays almost exclusively into a \Wbos and a \bquark. The properties of the top decay vertex $Wtb$ are determined by the structure of the weak interaction. In the Standard Model (SM) this interaction has a ($V - A$) structure, where $V$ and $A$ refer to the vector and axial vector components of the weak coupling. The \Wbos, which is produced as a real particle in the decay of top quarks, possesses a polarisation which can be left-handed, right-handed or longitudinal. The corresponding fractions, referred to as helicity fractions, are determined by the $Wtb$ vertex structure and the masses of the particles involved. Calculations at next-to-next-to-leading order (NNLO) in QCD predict the fractions to be $\FL = 0.311 \pm 0.005$, $\FR = 0.0017 \pm 0.0001$, $\FN = 0.687 \pm 0.005$~\cite{WHel_theo}. 

By measuring the polarisation of the \Wbos\ with high precision, the SM prediction can be tested, and new physics processes which modify the structure of the $Wtb$ vertex can be probed. 
The structure of the $Wtb$ vertex can be expressed in a general form using left- and right-handed vector ($V_{\mathrm{L/R}}$) and tensor ($g_{\mathrm{L/R}}$) couplings:
\begin{eqnarray}
\mathcal{L}_{Wtb} & = & - \frac{g}{\sqrt 2} \bar b \, \gamma^{\mu} \left( V_{\mathrm{L}} P_{\mathrm{L}}
  + V_{\mathrm{R}} P_{\mathrm{R}} \right) t\; W_\mu^-  - \frac{g}{\sqrt 2} \bar b \, \frac{i \sigma^{\mu \nu} q_\nu}{m_W}
  \left( g_{\mathrm{L}} P_{\mathrm{L}} + g_{\mathrm{R}} P_{\mathrm{R}} \right) t\; W_\mu^- + \mathrm{h.c.} 
\label{eq:Wtb}
\end{eqnarray}

Here, $P_{\mathrm{L/R}}$ refer to the left- and right-handed chirality projection operators, $m_W$ to the \Wbos\ mass, and $g$ to the weak coupling constant. At tree level, all of the vector and tensor couplings vanish in the SM, except $V_{\mathrm{L}}$, which corresponds to the CKM matrix element $V_{tb}$ and has a value of approximately one. Dimension-six operators, introduced in effective field theories, can lead to anomalous couplings, represented by non-vanishing values of $V_{\mathrm{R}}$, $g_{\mathrm{L}}$ and $g_{\mathrm{R}}$~\cite{Buchmuller:1985jz,AguilarSaavedra:2008zc,Zhang:2010dr}.

The \Wbos\ helicity fractions can be accessed via angular distributions of polarisation analysers. Such analysers are \Wbos\ decay products whose angular distribution is sensitive to the \Wboson\ polarisation and determined by the $Wtb$ vertex structure. In case of a leptonic decay of the \Wbos\ ($W\rightarrow \ell \nu$), the charged lepton serves as an ideal analyser: its reconstruction efficiency is very high and the sensitivity of its angular distribution to the \Wboson\ boson polarisation is maximal due to its weak isospin component $T_3 = -\frac{1}{2}$. If the \Wbos\ decays hadronically  ($W\rightarrow q\bar{q}^{\prime}$), the \dQ is used, as it carries the same weak isospin as the charged lepton. This provides it with the same analysing power as the charged lepton, which is only degraded by the lower reconstruction efficiency and resolution of jets compared to charged leptons. 
The reconstruction of the \dQ is in particular difficult as the two decay products of a hadronically decaying \Wbos are experimentally hard to separate.  In the \Wbos\ rest frame, the differential cross-section of the analyser follows the distribution
\begin{equation}
\frac{1}{\sigma} \frac{{\textrm d} \sigma}{{\textrm d} \costheta} = \frac{3}{4} \left( 1-\cos^{2} \theta^{*} \right) \, \FN + \frac{3}{8} \left(1-\costheta \right)^2 \, \FL + \frac{3}{8} \left(1 + \costheta\right)^{2} \, \FR \,,
\label{eq:dist}
\end{equation}
which directly relates the \Wbos\ helicity fractions \Fi\ to the angle $\theta^{*}$ between the analyser and the reversed direction of flight of the \bquark\ from the top quark decay in the $W$ boson rest frame. 
Previous measurements of the \Wbos\ helicity fractions from the ATLAS, CDF, CMS and D0 collaborations show 
agreement with the SM within the 
uncertainties~\cite{Aaltonen:2012rz,Aaltonen:2010ha,Abazov:2010jn,TOPQ-2011-10,CMS-TOP-13-008}. 

In this paper, the \Wbos\ helicity fractions are measured in top quark pair (\ttbar) events. Data corresponding to 
an integrated luminosity of 20.2~\ifb of proton--proton ($pp$) collisions, produced at the LHC with a \com\ energy 
of \rts = 8 \tev, and recorded with the ATLAS~\cite{PERF-2007-01} detector, are analysed. The final state of the \ttbar\ 
events is characterised by the decay of the \Wboson\ bosons. This analysis considers the \ljets\ channel in which 
one of the \Wboson\ bosons decays leptonically and the other decays hadronically. Both \Wbos\ decay modes are utilised for the measurement of \costheta. The signal 
selection and reconstruction includes direct decays of the \Wbos\ into an electron or muon as well as \Wbos\ decays 
into a $\tau$-lepton which subsequently decays leptonically.

\section{The ATLAS detector}
\label{sec:atlas}

The ATLAS experiment at the LHC is a multi-purpose particle detector with a forward-backward symmetric cylindrical geometry and a near $4\pi$ coverage in solid angle.\footnote{\AtlasCoordFootnote}
It consists of an inner tracking detector surrounded by a thin superconducting solenoid providing a \SI{2}{\tesla} axial magnetic field, electromagnetic and hadron calorimeters, and a muon spectrometer. The inner tracking detector covers the pseudorapidity range $|\eta| < 2.5$. It consists of silicon pixel, silicon microstrip, and transition-radiation tracking detectors. Lead/liquid-argon (LAr) sampling calorimeters provide electromagnetic energy measurements with high granularity. A hadron (steel/scintillator-tile) calorimeter covers the central pseudorapidity range ($|\eta| < 1.7$). The end-cap and forward regions are instrumented with LAr calorimeters for electromagnetic and hadronic energy measurements up to $|\eta| = 4.9$. The muon spectrometer surrounds the calorimeters and is based on three large air-core toroid superconducting magnets with eight coils each. Its bending power ranges from \num{2.0} to \SI{7.5}{\tesla\metre}. It includes a system of precision tracking chambers and fast detectors for triggering. A three-level trigger system is used to select events. The first-level trigger is implemented in hardware and uses a subset of the detector information to reduce the accepted rate to at most \SI{75}{\kilo\hertz}. This is followed by the high-level trigger, two software-based trigger levels that together reduce the accepted event rate to \SI{400}{\hertz} on average depending on the data-taking conditions.

\section{Data and simulated samples}
\label{sec:dataAndMCsamples}

The data set consists of $pp$ collisions, recorded at the 
LHC with \rts = 8 \tev, and corresponds to an integrated luminosity of 
$20.2~\ifb$. 
Single-lepton 
triggers with a threshold of 24 \gev\ of transverse momentum (energy) for isolated 
muons (electrons) and 36 (60) \gev\ for muons (electrons) without an isolation criterion are 
used to select \ttbar\ candidate events. 
The lower trigger thresholds 
include isolation requirements on the candidate lepton, resulting in 
inefficiencies at high \pT that are recovered by the triggers with 
higher \pT thresholds.

Samples obtained from Monte Carlo (MC) simulations are used to characterise the detector response and reconstruction efficiency of \ttbar\ events,
estimate systematic uncertainties and predict the background contributions from various processes. The
response of the full ATLAS detector is simulated~\cite{SOFT-2010-01} using 
\GEANT4~\cite{SAMPLES-G4}. For the
estimation of some systematic uncertainties, generated samples are passed through a faster simulation with parameterised showers in the calorimeters~\cite{ATL-PHYS-PUB-2010-013}, while still using the full simulation of the tracking systems.
Simulated events include the effect of multiple $pp$ collisions from the same and nearby bunch-crossings
(in-time and out-of-time pile-up) and are reweighted to match the number of collisions observed
in data. 
All simulated samples are normalised using the most precise cross-section calculations available.

Signal \ttbar\ events are generated using the next-to-leading-order (NLO) QCD 
MC event generator \POWHEGBOX~\cite{powheg1,powheg2,powheg3, powheg4} 
using the {\textsc CT10} parton distribution function (PDF) set~\cite{CT10}. 
\POWHEGBOX\ is interfaced to \PYTHIA\ 6.425~\cite{pythia} (referred to as the \powpy\ sample), which is used to model the showering and hadronisation, with 
the {\textsc CTEQ6L1} PDF set~\cite{cteq6l} and 
a set of tuned parameters called the {\textsc Perugia2011C} tune~\cite{perugia_tune} for the modelling of the 
underlying event. The model parameter $h_{\text{damp}}$ is set to \mtop and controls 
matrix element to parton shower matching in \POWHEGBOX\ and 
effectively regulates the amount of high-\pt\ radiation.

The \ttbar\ cross-section is $\sigma(t\bar{t}) = 253^{+13}_{-15}$~pb. This 
value is the result of a NNLO QCD calculation 
that includes resummation of next-to-next-to-leading logarithmic soft gluon terms with 
top++2.0~\cite{Cacciari:2011hy, Beneke:2011mq, Baernreuther:2012ws, Czakon:2012pz, Czakon:2012zr, Czakon:2013goa, Czakon:2011xx}.

A sample generated with \POWHEGBOX\ interfaced with 
\HERWIG~6.520~\cite{herwig} using {\textsc Jimmy} 4.31~\cite{jimmy} to simulate the underlying event (referred to as the \powhw\ sample) is compared to a \powpy\ sample to assess the impact of the different parton shower models. 
For both the \powhw\ sample and this alternate \powpy\ sample, the $h_{\text{damp}}$ parameter is set to infinity.

To estimate the uncertainty due 
to the choice of the MC event generator, an alternate \ttbar\ MC sample is produced with MC@NLO~\cite{Frixione:2002ik, Frixione:2003ei} 
with the {\textsc CT10} PDF set interfaced to \HERWIG~6.520 using the AUET2 tune~\cite{ATL-PHYS-PUB-2011-008} and the {\textsc CT10} PDF set for showering and hadronisation.
In addition, samples generated with \POWHEGBOX\ interfaced 
to \PYTHIA with variations in the amount of QCD initial- and final-state radiation (ISR/FSR) 
are used to estimate the effect of such uncertainty. The factorisation and renormalisation scales and the $h_{\text{damp}}$ parameter in \POWHEGBOX\ as well as the  
 transverse momentum scale of the space-like parton-shower evolution in \PYTHIA\ are varied within the constraints obtained from an ATLAS measurement of \ttbar\ production in association with jets~\cite{TOPQ-2011-21}.

Single-top-quark-processes for the $t$-channel, $s$-channel and $Wt$ associated production are also simulated with \POWHEGBOX~\cite{powheg_s_and_t, powheg_Wt} using the CT10 PDF set. The samples are 
interfaced to \PYTHIA\ 6.425 with the CTEQ6L1 PDF set and the Perugia2011C underlying event tune. 
Overlaps between the \ttbar\ and $Wt$ final states are removed~\cite{Sgtop_WtAndTtbar}. The single-top-quark samples are 
normalised using the approximate NNLO theoretical cross-sections~\cite{stopxs,stopxs_2,stopxs_3} calculated with 
the MSTW2008 NNLO PDF set~\cite{mstw1,mstw2}. All \ttbar\ and single-top samples are generated assuming a top quark mass of 172.5\,\GeV,  compatible with the ATLAS measurement of $m_{t} = 172.84 \pm 0.70\,\GeV$~\cite{TOPQ-2016-03}.

Events with a \Wboson or \Zboson~boson produced in association with jets are generated using the leading-order (LO) 
event generator \ALPGEN 2.14~\cite{alpgen} with up to five additional partons and the CTEQ6L1 PDF set, interfaced 
to \PYTHIA~6.425 for the parton showering and hadronisation. 
Separate samples for $W/Z$+light-jets, $W/Zb\bar{b}$+jets, $W/Zc\bar{c}$+jets 
and $Wc$+jets were generated. A parton--jet matching scheme (``MLM matching'') \cite{mlm} is employed to avoid 
double-counting of jets generated from the matrix element and the parton shower. Overlap between 
the $W/ZQ\bar{Q} \left(Q=b,c \right)$ events generated at the matrix element level and those generated 
by the parton shower evolution of the $W/Z$+light-jets sample are removed with an angular separation algorithm. If the 
angular distance \dR\ between the heavy-quark pair is 
larger than 0.4, the matrix element prediction is used instead of the parton shower prediction.
Event yields from the $Z$+jets background are normalised using their inclusive NNLO theoretical cross-sections~\cite{VjetsNorm}. The predictions of normalisation and flavour composition of the \wjets\ background are affected
by large uncertainties. Hence, a data-driven technique is used to determine both the inclusive normalisation and the heavy-flavour fractions of this process. The approach followed exploits the fact that the \Wpm\ boson production is charge-asymmetric at 
a $pp$~collider. 
The $W$ boson charge asymmetry depends on the flavour composition of the sample. 
Thus, correction factors estimated from data are used to rescale the fractions of  
$Wb\bar{b}/c\bar{c}$+jets, $Wc$+jets and $W+$light-jets events in the MC simulation: 
$K_{bb}$ = $K_{cc}$ = 1.50 $\pm$ 0.11~\text{(stat. + syst.)},  $K_{c}$ = 1.07 $\pm$ 0.27~\text{(stat. + syst.)} 
and $K_{\text{light}}$ = 0.80 $\pm$ 0.04~\text{(stat. + syst.)}~\cite{TOPQ-2014-16}.

Diboson samples ($WW$, $ZZ$, $WZ$) are generated using the \textsc{Sherpa} 1.4.1~\cite{Sherpa} event generator
with the \textsc{CT10} PDF set, with massive $b$- and
$c$-quarks and with up to three additional partons in the LO matrix elements. The yields 
of these backgrounds are normalised using their NLO QCD theoretical
cross-sections~\cite{Campbell:1999ah}.

Multijet events can contain jets misidentified as leptons or non-prompt leptons from hadron decays and hence satisfy the selection criteria of the \ljets\ topology. This source of background events is referred to as fake-lepton background and is estimated using a data-driven approach (``matrix method'') which is based on the measurement of lepton selection efficiencies using different identification and isolation criteria~\cite{ATLAS-CONF-2014-058}.

\section{Event selection and \ttbar\ reconstruction}
\label{sec:selection}

\subsection*{Object reconstruction}
The final state contains electrons, muons, jets with some of them originating from $b$-quarks, as well as missing transverse momentum.

Electrons are reconstructed from energy depositions in the electromagnetic calorimeter matching tracks in the inner detector. The transverse component of the energy deposition has to exceed 25\,\gev\ and the pseudorapidity of the energy cluster, \etaclus, has to fullfil $\absetaclus < 2.47$, excluding the transition region between the barrel and end-cap sections of the electromagnetic calorimeter at $1.37 < \absetaclus\ < 1.52$. Electrons are further required to have a longitudinal impact parameter with respect to the hard-scattering vertex of less than 2\,mm. 

To reduce the background from non-prompt electrons (i.e. electrons produced within jets), electron candidates are also required to be isolated. Two $\eta$-dependent isolation criteria are applied. The first one considers the energy deposited in the calorimeter cells within a cone of size $\Delta R = 0.2$ around the electron direction. The second one sums the transverse momenta (\pt) of all tracks with \pt\ > 400 \mev\ within a cone of size $\Delta R = 0.3$ around the electron track. For each quantity, the transverse energy or momentum of the electron are subtracted. 
The isolation requirement is applied in such a way as to retain 90\% of signal electrons, independent of their \pt\ value. This constant efficiency is verified in a data sample of  \Zee decays~\cite{ATLAS-CONF-2014-032}.

For the reconstruction of muons, information from the muon spectrometer and the inner detector is combined. The combined muon track must satisfy $\pt > 25\,\gev$ and $\abseta < 2.5$. The longitudinal impact parameter with respect to the hard-scattering vertex (defined in next section) is required to be less than 2\,mm. 
Furthermore, muons are required to satisfy a $\pt$-dependent
track-based isolation requirement.  The scalar sum of the track $\pt$ in a cone of
variable size $\Delta R < 10\,\gev/\pt^\mu$ around the muon
(excluding the muon track itself) has to be less than 5\,\% of the muon $\pt$.

Jets are reconstructed from topological clusters \cite{PERF-2007-01} built from energy depositions in the calorimeters using 
the \antikt algorithm \cite{antikt, fastjet} with a radius parameter of 0.4. Before being processed by the jet-finding 
algorithm, the topological cluster energies are corrected using a local calibration 
scheme~\cite{ATL-LARG-PUB-2008-002, Cojocaru:2004jk} to account for inactive detector material, out-of-cluster leakage and the noncompensating calorimeter response. After energy calibration \cite{PERF-2012-01}, the jets are required to have $\pt > 25\,\gev$ and $\abseta < 2.5$. To suppress jets from pile-up, the jet vertex fraction\footnote{The jet vertex fraction is defined as the scalar sum of the transverse momenta of a jet's tracks stemming from the primary collision vertex divided by the scalar sum of the transverse momenta of all tracks in a jet.} is required to be above 0.5 for all jets with \pt~<~50~\gev\ and \abseta~<~2.4. As all electron candidates are also reconstructed as jets, the  closest jet within a cone of size \dR~=~0.2 around an electron candidate is discarded to avoid double-counting of electrons as jets.  After this removal procedure, electrons within \dR~=~0.4 of any remaining jet are removed.

Jets are identified as originating from the hadronisation of a \bquark\ ($b$-tagged) via a multivariate algorithm ~\cite{PERF-2012-04}. It makes use of the lifetime and mass of $b$-hadrons and accounts for displaced tracks and topological properties of the jets. A working point with 70\,\% efficiency to tag a \bquark\ jet (\bt-jet) is used. The rejection factor for light-quark and gluon jets (light jets) is around 130 and about 5
for charm jets, as determined for $b$-tagged jets with
\pt~>~20~\gev\ and $|\eta|<2.5$ in simulated \ttbar events. The simulated $b$-tagging efficiency is corrected to that measured in data using calibrations from statistically independent event samples of \ttbar\ pairs decaying into a $b\bar{b}\ell^{+}\ell^{-}\nu_{\ell}\bar{\nu_{\ell}}$ final state~\cite{ATLAS-CONF-2014-004}. 

The reconstruction of the transverse momentum of the neutrino from the leptonically decaying \Wbos\ is based on the negative vector sum of all energy deposits and momenta of reconstructed and calibrated objects in the transverse plane (missing transverse momentum with magnitude \met) as well as unassociated energy depositions~\cite{PERF-2011-07}.

\subsection*{Event selection}
Events are selected from data taken in stable beam conditions with all relevant detector components being functional. 
At least one primary collision vertex is required with at least five associated tracks with \pt~>~400~\mev. If more than one primary vertex is reconstructed, the one with the largest scalar sum of transverse momenta is selected as the hard-scattering vertex. If the event contains at least one jet with $\pt > 20\,\gev$ that is identified as out-of-time activity from a previous $pp$ collision or as calorimeter noise~\cite{ATLAS-CONF-2012-020}, the event is rejected.

In order to select events from \ttbar decays in the \ljets\ channel, exactly one reconstructed electron or muon with \pt$>25$ \gev\ and at least four jets, of which at least one is $b$-tagged, are required. A match ($\Delta R < 0.15$) between the offline reconstructed electron or muon and the lepton reconstructed by the high-level trigger is required. The selected events are separated into two orthogonal $b$-tag regions: one region with exactly one $b$-tag and a second region with two or more $b$-tags. Thus, the data sample is split into four channels depending on the lepton flavour and the $b$-jet multiplicity: ``$e$+jets, 1 $b$-tag'', ``$e$+jets, $\geq$2 $b$-tags'', ``$\mu$+jets, 1 $b$-tag'' and ``$\mu$+jets, $\geq$2 $b$-tags''.

For events with one \bt-tag, \met is required to be greater than 20~\gev\ and the sum of \met and transverse mass of the leptonically decaying \Wbos, $\mtw$, is required to be greater than 60~\gev\ in order to suppress multijet background. In the case of two \bt-tags, no further requirement on the \met and transverse mass of the \Wbos is applied.

After this selection, the \ttbar\ candidate events are reconstructed using a kinematic likelihood fit as described next.
\subsection*{Reconstruction of the \ttbar\ system}
\label{sec:reconstruction}
The measurement of the \Wbos polarisation in \ttbar\ events requires the reconstruction and identification of all \ttbar\ decay products. For this, a kinematic likelihood fitter (KLFitter)~\cite{klfitter} is utilised. It maps the four model partons (two $b$-quarks and the $q\bar{q}^{\prime}$ pair from a \Wboson\ boson decay) to four reconstructed jets. The numbers of jets used as input for KLFitter can be larger than four. The two jets with the largest output of the $b$-tagging algorithm together with two (three) remaining jets with the highest \pt\ were chosen as KLFitter input as this selection leads to the highest reconstruction efficiency for events with four (at least five) jets. For each of the $4! = 24$ ($5!=120$ for events with at least five jets) possible jet-to-parton permutations, it maximises a likelihood, $\mathscr{L}$, that incorporates Breit--Wigner distributions for the \Wbos\ and top quark masses as well as transfer functions mapping the reconstructed jet and lepton energies to parton level or true lepton level, respectively. The expression for the likelihood is given by
\begin{eqnarray}
&& \mathscr{L}=BW(m_{q_1 q_2 q_3}| m_{t}, \Gamma_{t})\cdot BW(m_{q_1 q_2}| \mW, \Gamma_{W})\cdot BW(m_{q_4 \ell \nu}| m_{t}, \Gamma_{t})\cdot BW(m_{\ell \nu} | \mW, \Gamma_{W}) \nonumber \\
&& \cdot W(E_{\text{jet}_1}^{\text{meas}}|E_{q_1})\cdot W(E_{\text{jet}_2}^{\text{meas}}|E_{q_2})\cdot W(E_{\text{jet}_3}^{\text{meas}}|E_{q_3})\cdot W(E_{\text{jet}_4}^{\text{meas}}|E_{q_4})\nonumber \\ && \cdot W(E_{\ell}^{\text{meas}}|E_{\ell})\cdot W(E^{\text{miss,} x}|p^{x}_{\nu})\cdot W(E^{\text{miss,} y}|p^{y}_{\nu}) .
\label{eq:KLF}
\end{eqnarray}
where the $BW(m_{ij(k)}| m_{t/W}\Gamma_{t/W})$ terms are the Breit-Wigner functions used to evaluate the mass of composite reconstructed particles (\Wboson\ bosons and top quarks) and $W(E_{i}^{\text{meas}}|E_{j})$ are the transfer functions, with $E_{i}^{\text{meas}}$ being the measured energy of object $i$ and $E_{j}$ the ``true'' energy of the reconstructed parton $j$ or true lepton~$\ell$. The transverse components $p^{x/y}_{\nu}$ of the neutrino momentum are mapped to the missing transverse momentum $E^{\text{miss,} x/y}$ via transfer functions $W(E^{\text{miss,} x/y}|p^{x/y}_{\nu})$. Individual transfer functions for electrons, muons, \bt-jets, light jets (including $c$-jets) and missing transverse momentum are used. These transfer functions are obtained from \ttbar\ events  simulated with \MCatNLO. The top quark decay products are uniquely matched to reconstructed objects to obtain a continuous function describing the relative energy difference between parton and reconstructed level as a function of the parton-level energy. Individual parameterisations are derived for different regions of \abseta. The measurement of the \Wbos\ polarisation in the \ljets\ channel is performed for both the top and the anti-top quarks in each event. The anti-down-type quark from the top quark decay (down-type quark from the anti-top quark decay) is used as the hadronic analyser and the charged lepton from the decay of the anti-top quark (charged anti-lepton from the top quark decay) as the leptonic analyser. 

Since the likelihood defined in Equation~\eqref{eq:KLF} is invariant under exchange of the \Wboson\ decay products, it needs further extensions to incorporate information related to down-type quarks. This is achieved by multiplying the likelihood by probability distributions of the \bt-tagging algorithm output as a function of the transverse momentum of the jets. These probability distributions are obtained from \MCatNLO\ for $b$-quark jets as well as $u/c$- and $d/s$-quark jets. Since the \Wbos\ decays into a pair of charm and strange quarks in 50\,\% of decays into hadrons, the higher values of the \bt-tagging algorithm output for the charm quark allows for a separation of the two. This increases the fraction of events with correct matching of the two jets originating from a \Wboson\ boson decay to the corresponding up- and down-quark type jet to 60\,\%, compared to 50\,\% for the case of no separation power. The extended likelihood is normalised with respect to the sum of the extended likelihoods for all 120 (24) permutations and this quantity is called the ``event probability''. This up- versus down-type quark separation method was established in an ATLAS measurement of the \ttbar\ spin correlation in the \ljets\ channel~\cite{TOPQ-2013-01}.

\begin{figure}[!htbp]
  \begin{center}
    \mbox{
      \subfloat[]{\includegraphics[width=0.47\textwidth]{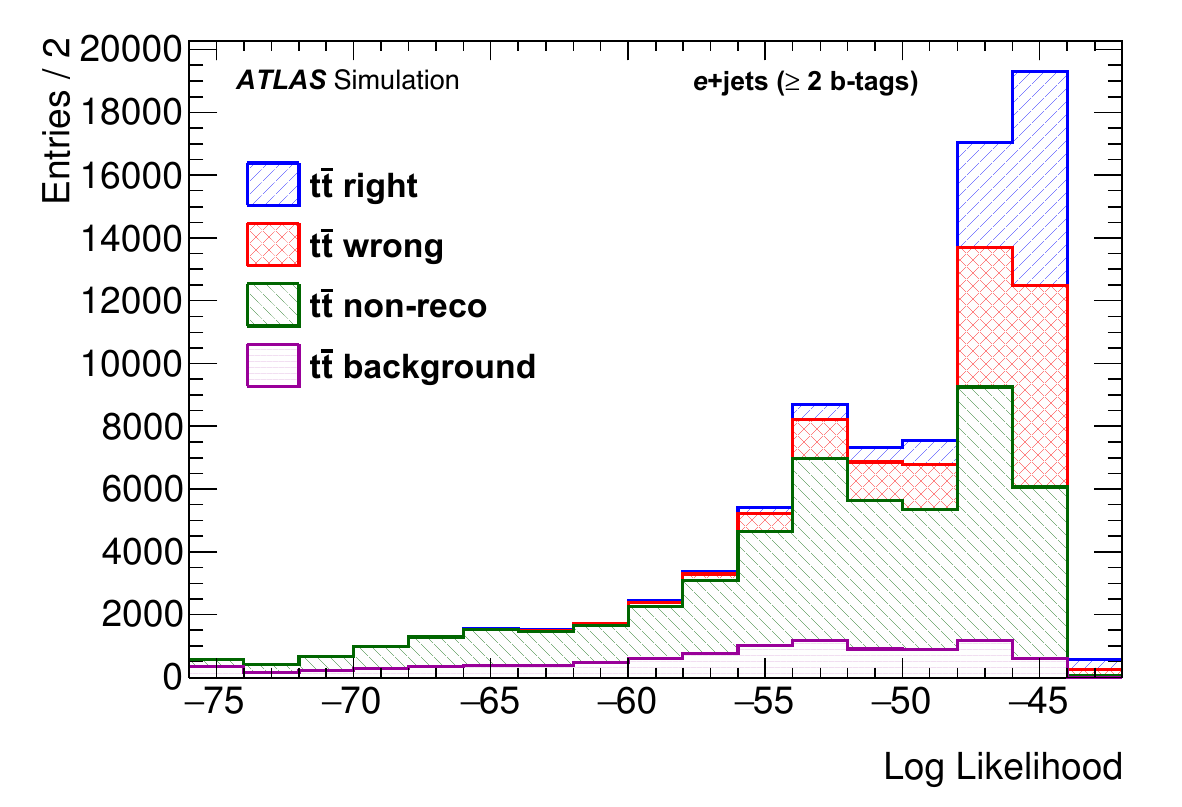}\label{fig:lh_stack}}    \quad 
      \subfloat[]{\includegraphics[width=0.47\textwidth]{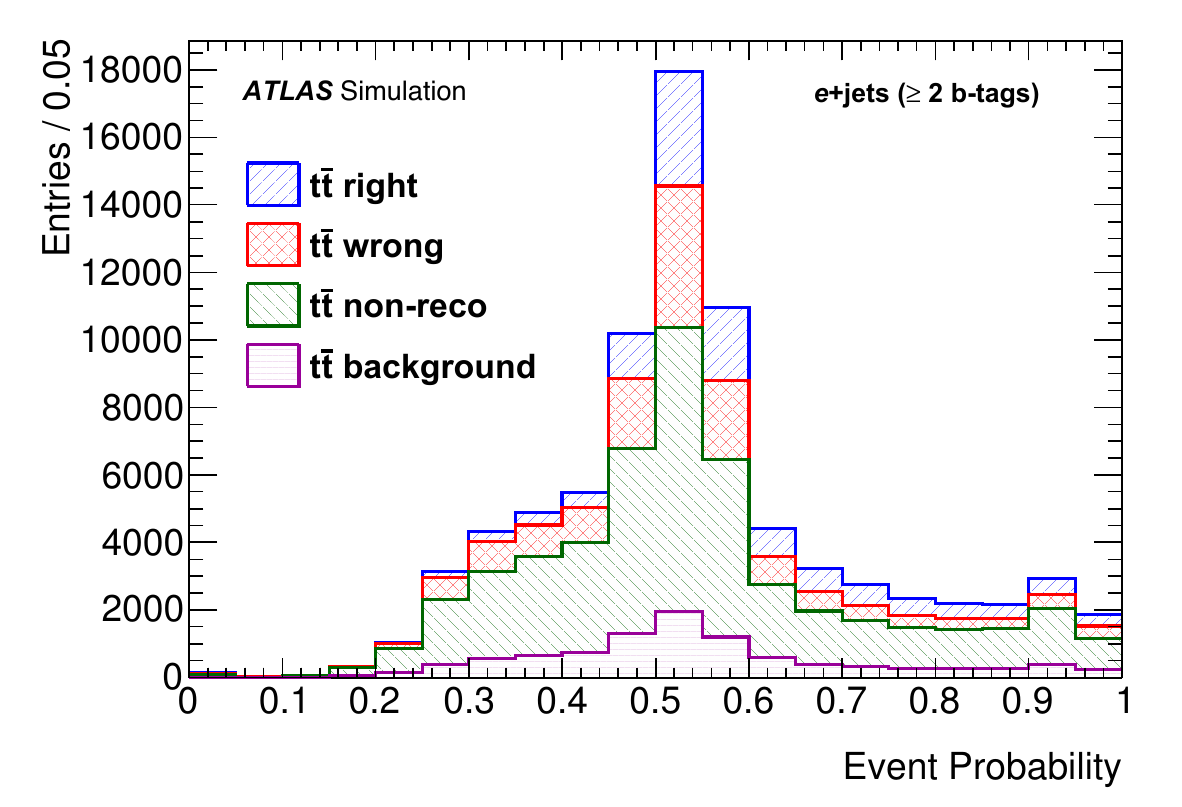}\label{fig:evProb_stack}}    \quad 
    }
    \caption{\protect\subref{fig:lh_stack} Logarithm of the likelihood value as output for reconstructed \ttbar\ events of the selected (best) jet-to-parton permutation. \protect\subref{fig:evProb_stack} Event probability for the selected (best) jet-to-parton permutation. Both distributions show events in the \ejets\ channel with $\geq 2$ \bt-tags. Events from a \ttbar\ signal sample are split into events where the \ttbar\ pairs do not decay via the \ljets\ channel (``\ttbar\ background''), events where not all \ttbar\ decay products have been reconstructed (``\ttbar\ non-reco''), as well as correctly (``\ttbar\ right'') and incorrectly (``\ttbar\ wrong'') reconstructed \ttbar\ systems.}
    \label{fig:logLH}
  \end{center}
\end{figure}

The permutation with the largest event probability is chosen. Figure~\ref{fig:lh_stack} shows the distributions of the logarithm of the likelihood value for the permutation with the highest event probability for simulated \ttbar\ events. Correctly reconstructed events (``\ttbar\ right'') peak at high values of the likelihood. Other contributions come from incorrect assignments of jets (i.e. choosing the wrong permutation, ``\ttbar\ wrong''), non-reconstructable events where for example a quark is out of the acceptance (``\ttbar\ non-reco'') and \ttbar\ events which do not have a \ljets\ topology (such as dileptonic \ttbar\ events, ``\ttbar\ background''). In Figure~\ref{fig:evProb_stack} the corresponding distribution of the event probability is shown. The peak at 0.5 corresponds to events where no separation between up- and down-type quarks is achieved, leading to two permutations with similar event probabilities. High event probability indicates a correct down-type quark reconstruction. 

To select the final data sample, the event probability is used to obtain the best jet-to-parton permutation per event. Events are required to have a reconstruction likelihood of $\log{\mathscr{L}} > -48$ to reject poorly reconstructed \ttbar\ events. The value of  $\log{\mathscr{L}} > -48$ was selected to minimise the expected statistical uncertainty. The fraction of events where all jets were correctly assigned to the corresponding partons out of all events that have the corresponding jets present varies between 45-50\,\%.  
The event yields after the final event selection are presented in Table~\ref{tab:eventyields_selection}.

\begin{table}
\centering 
\begin{tabular}{l|S[separate-uncertainty]|S[separate-uncertainty]|S[separate-uncertainty]|S[separate-uncertainty]}

\hline
\hline
 & \multicolumn{2}{c|}{\ejets} & \multicolumn{2}{c}{\mujets} \\\hline
Sample & {1 \bt-tag} & {$\geq2$ \bt-tags} & {1 \bt-tag} & {$\geq2$ \bt-tags} \\
\hline
\ttbar     		& 36500 	\pm 2300 	& 36000 	\pm 2300	& 43600 	\pm 2800 	& 42600 	\pm 2700 \\
Single top 	& 2000 	\pm 340 		& 974 	\pm 170 		& 2328 	\pm 400 		& 1102 	\pm 190\\
$W$+light-jets 	& 600 	\pm 30 		& 24 		\pm 1 		& 761 	\pm 38 		& 45 		\pm 2\\
$W+c$ 		& 1210 	\pm 300 	& 54 		\pm 13 		& 1440 	\pm 360 	& 51 		\pm 13\\
$W+bb/cc$ 	& 2730 	\pm 190 	& 538 	\pm 38 		& 3520 	\pm 250 	& 780 	\pm 55\\
$Z$+jets 		& 1200  	\pm 580 	& 330 	\pm 160 	& 610  	\pm 290 	& 158 	\pm 76\\
Diboson    	& 220 	\pm 100 	& 33  	\pm 16 		& 210 	\pm 100 	& 37  	\pm 18\\
Fake lepton     	& 2270 	\pm 680 	& 450 	\pm 130	& 1750 	\pm 520 	& 323 	\pm 97 \\
\hline
Total expected  & 46700 \pm 2600		& 38400 \pm 2300 	& 54200 	\pm 2900	& 45100 \pm 2800\\
\hline
Data       		& 45246 		 		& 40045 				& 53747 				& 46048  \\
\hline\hline
\end{tabular}
\caption{Expected and observed event yields in the four channels (``$e$+jets, 1 $b$-tag'', ``$e$+jets, $\geq$2 $b$-tags'', ``$\mu$+jets, 1 $b$-tag'' and ``$\mu$+jets, $\geq$2 $b$-tags'') after the final event selection including the cut on the reconstruction likelihood. Uncertainties in the normalisation of each sample include systematic uncertainties for the data-driven backgrounds ($W$+jets and fake leptons) and theory uncertainties for the \ttbar\ signal and the other background sources.}
\label{tab:eventyields_selection}
\end{table}

Figure~\ref{fig:recoplots} shows the likelihood and the event probability as well as the reconstructed \costheta\ distribution after the final event selection. Good agreement between data and prediction is achieved.

\begin{figure}[!htbp]
  \begin{center}
  \mbox{
    \subfloat[]{\includegraphics[width=0.4\textwidth]{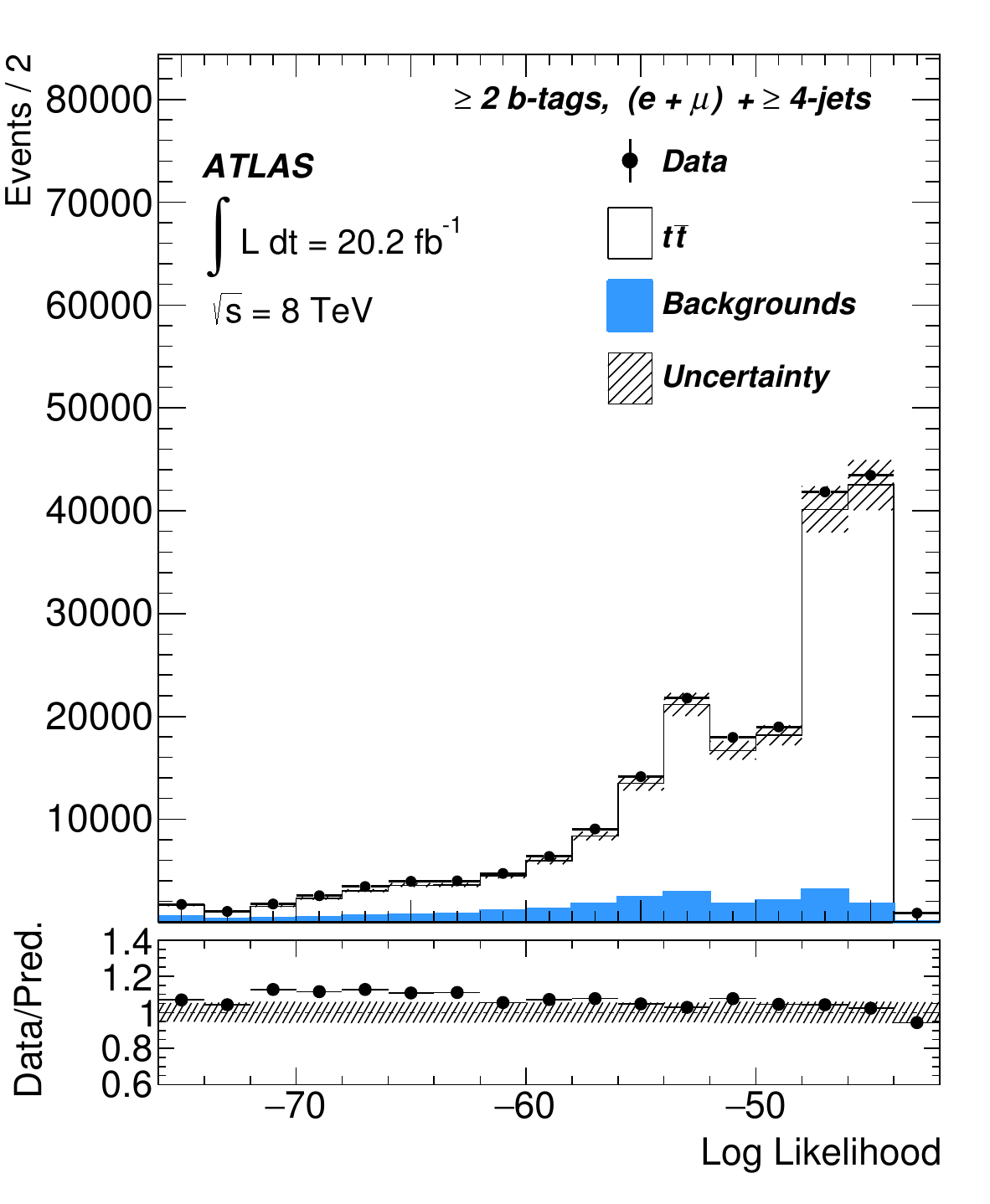}\label{fig:LogLikelihood_elmu}}\quad
    \subfloat[]{\includegraphics[width=0.4\textwidth]{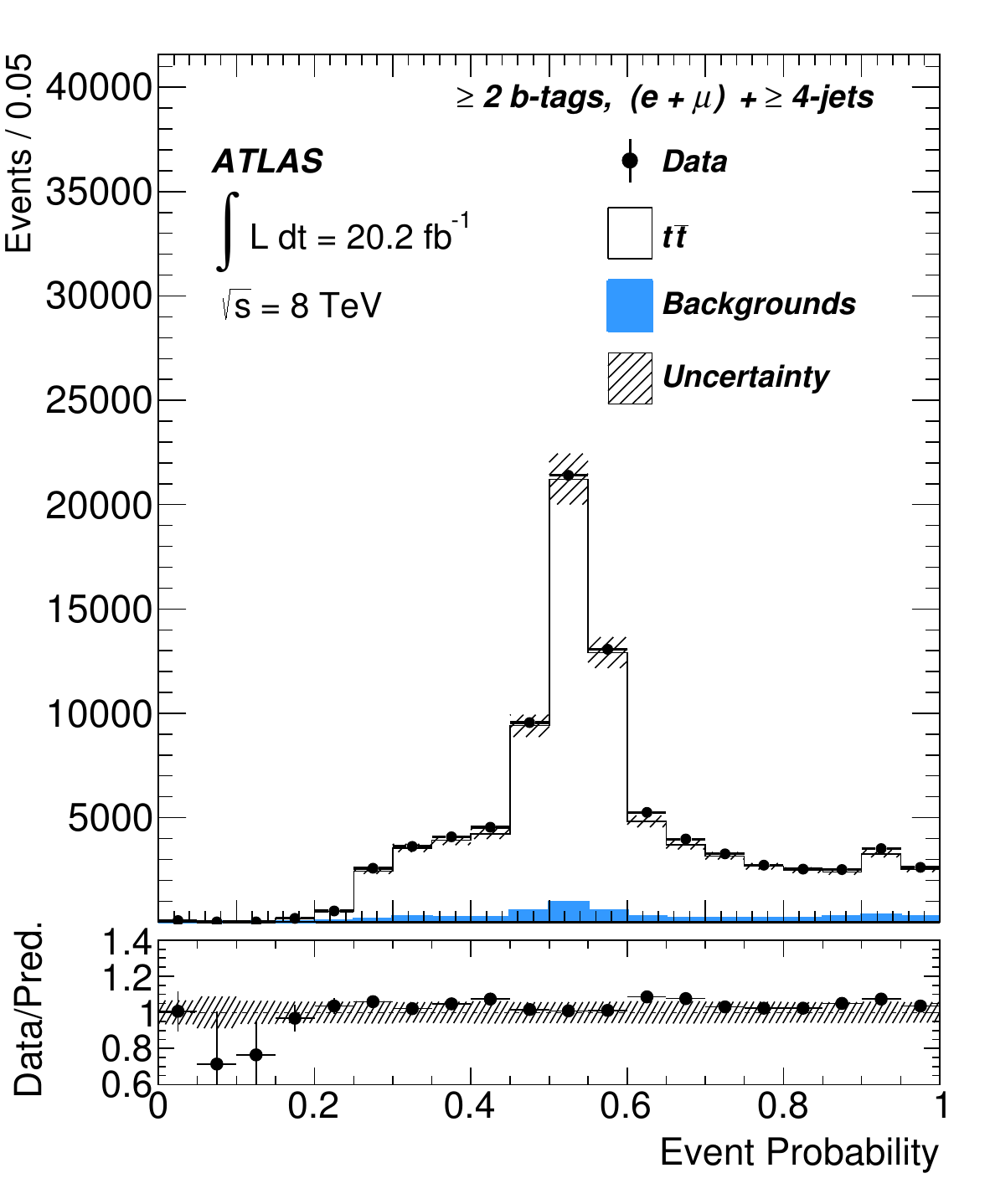}\label{fig:EventProbability_elmu}}\quad
  }
  \mbox{
    \subfloat[]{\includegraphics[width=0.4\textwidth]{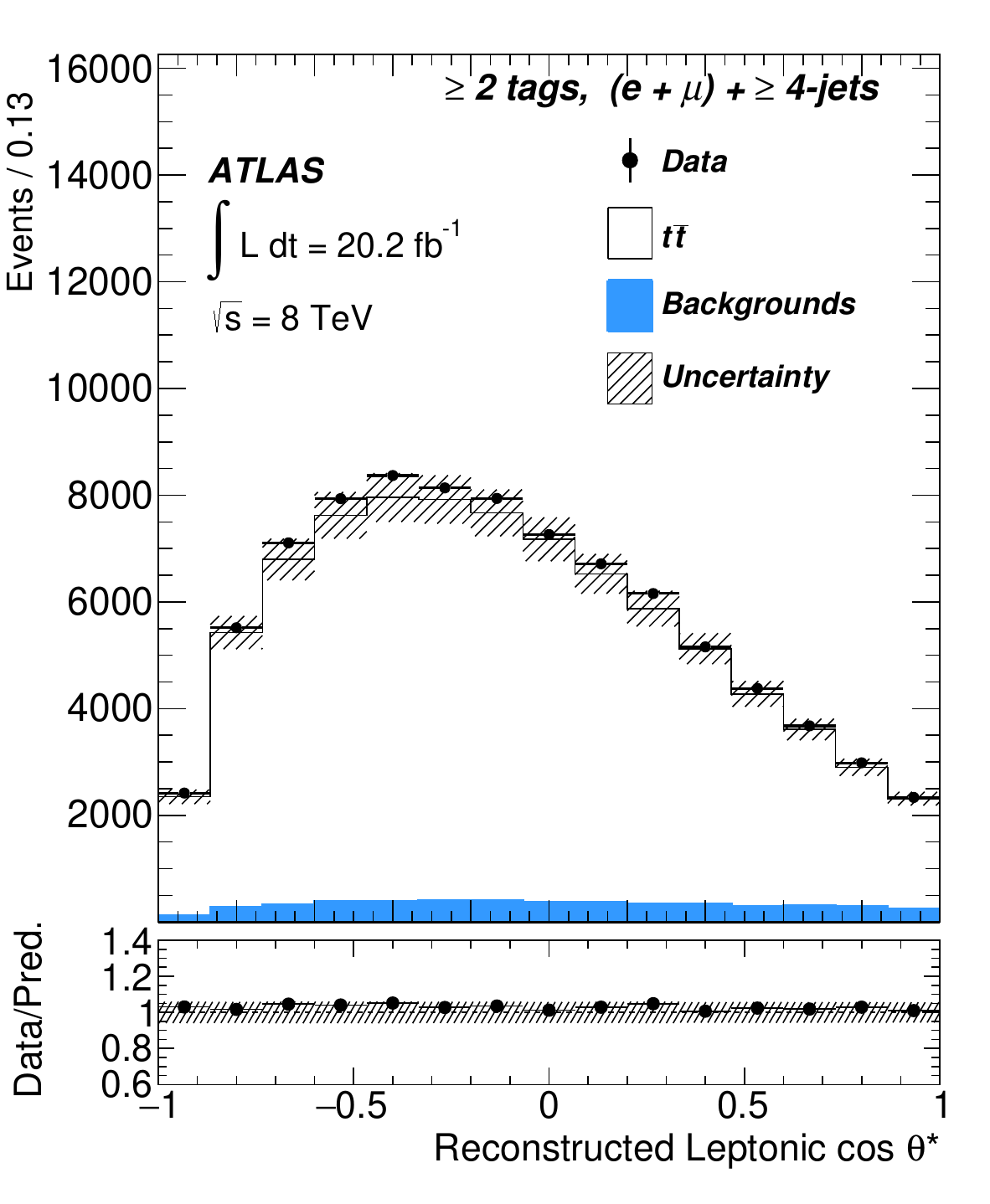}\label{fig:CosTheta_reco_lep_elmu}}\quad
    \subfloat[]{\includegraphics[width=0.4\textwidth]{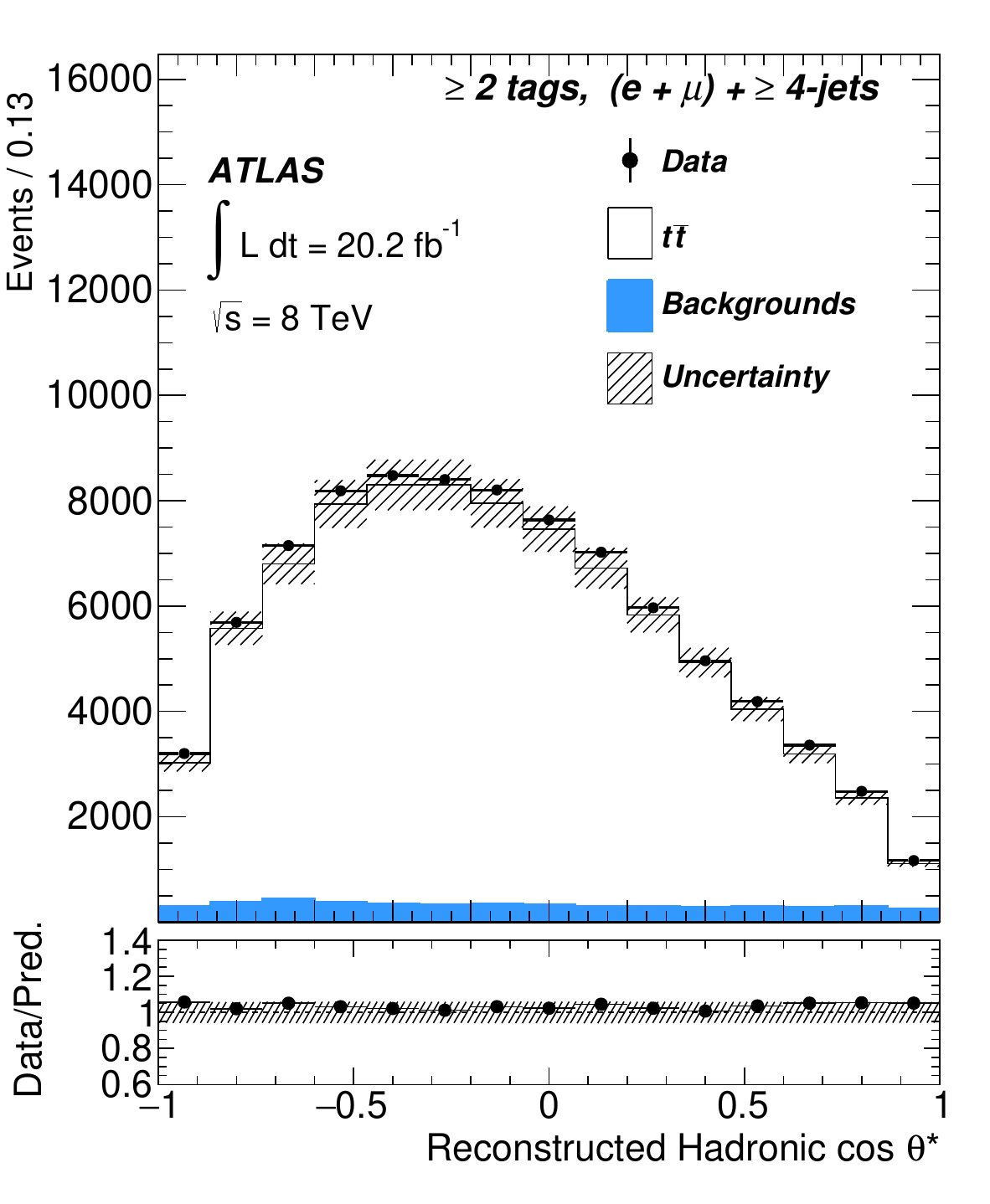}\label{fig:CosTheta_reco_had_elmu}}\quad
  }
  \caption{Measured and predicted distributions of \protect\subref{fig:LogLikelihood_elmu} likelihood and \protect\subref{fig:EventProbability_elmu} event probability from the kinematic fit and reconstructed \costheta\ distribution using \protect\subref{fig:CosTheta_reco_lep_elmu} the leptonic and \protect\subref{fig:CosTheta_reco_had_elmu} the hadronic analysers with $\geq$2 $b$-tags. The displayed uncertainties represent the Monte Carlo statistical uncertainty as well as the background normalisation uncertainties.}\label{fig:recoplots}
  \end{center}
\end{figure}

\section{Measurement of the $W$ boson helicity fractions}
\label{sec:analysis}

The $W$ boson helicity fractions \Fi\ are defined as the fraction of produced \ttbar\ events $N_i$ in a given polarisation state divided by all produced \ttbar\ events:
\begin{equation}
F_{\text{i}}= \frac{N_{{i}}}{N_{\mathrm{0}}+N_{\mathrm{L}}+N_{\mathrm{R}}}  \textrm{ \quad for } i\textrm{= 0, L, R}.
\label{eq:Frac}
\end{equation}
The selection efficiency $\epsilon_{i}^{\textrm{sel}}$ is different for each polarisation state and determines the number of selected events $n_i$:
\begin{equation}
n_i = \epsilon_{i}^{\textrm{sel}} N_{i}   \textrm{ \quad for } i\textrm{= 0, L, R}.
\label{eq:signal}
\end{equation}
Dedicated \ttbar\ signal templates for a specific \Fi\ are created by reweighting the simulated SM \ttbar\ events. These are produced by fitting the \costheta\ distribution for the full phase space and calculating per-event weights for each helicity fraction using the functional forms in Equation~\eqref{eq:dist}. Individual templates are created for each lepton flavour and $b$-tag channel. Figure \ref{fig:templates} shows the templates for the \mujets\ channel with $\geq$ 2 $b$-tags.

\begin{figure}[!htbp]
  \begin{center}
    \subfloat[]{\includegraphics[width=0.4\textwidth]{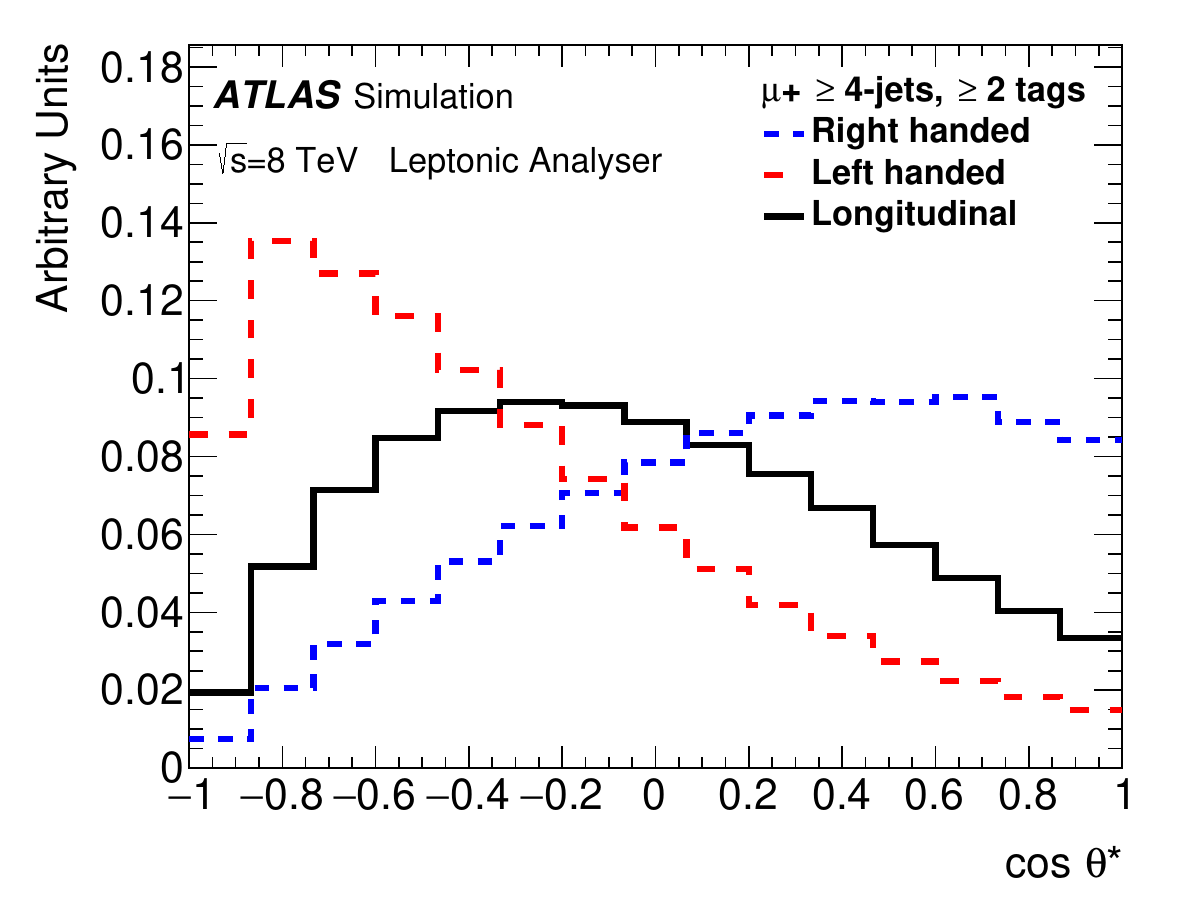}\label{fig:templates_lep}}\quad
    \subfloat[]{\includegraphics[width=0.4\textwidth]{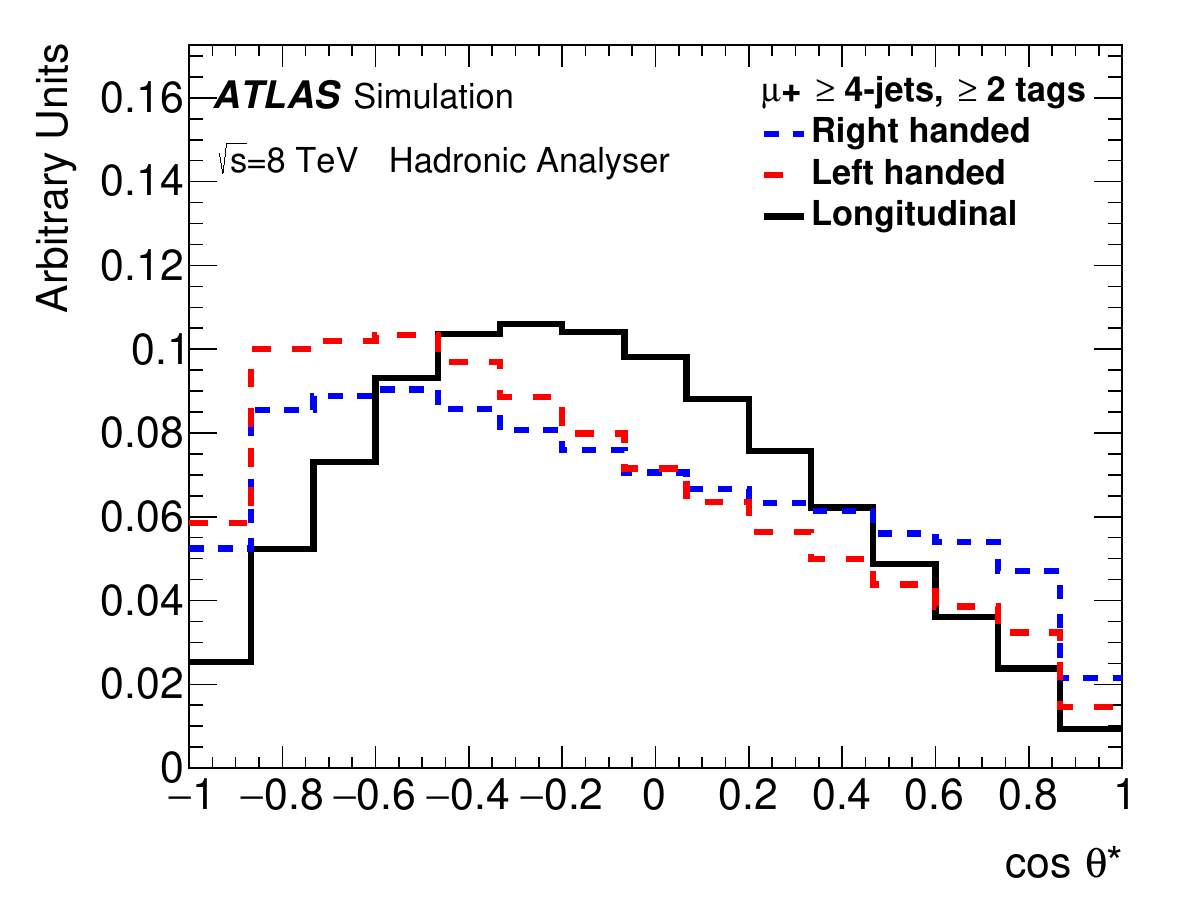}\label{fig:templates_had}}\quad
  \caption{Templates of the \costheta\ distributions for the individual helicity fractions in the \mujets\ channel with $\geq$ 2 $b$-tags for the \protect\subref{fig:templates_lep} leptonic and \protect\subref{fig:templates_had} hadronic analyser.}\label{fig:templates}
  \end{center}
\end{figure}

In addition to these signal templates, templates are derived for each source $j$ of background events. These are independent of the helicity fractions \Fi. Five different background templates are included: three $W$+jets templates ($W$+light-jets, $Wc$+jets and $Wc\bar{c}/b\bar{b}$+jets), a fake-lepton template, and one template for all remaining backgrounds, including contributions from electroweak processes (single top, diboson and $Z$+jets). The total number of expected events $n_{\textrm{exp}}$ in each channel is then given by
\begin{equation}
n_{\textrm{exp}}= n_{\textrm{0}} + n_{\textrm{L}} +n_{\textrm{R}}+n_{\textrm{$W$+light}}+n_{\textrm{$W$+$c$}}+n_{\textrm{$W$+$bb/cc$}}+n_{\textrm{fake}}+n_{\textrm{rem. bkg.}}.
\label{eq:background}
\end{equation}

The signal and background templates are used to perform a likelihood fit with the number of background events $n_{\textrm{bkg,}j}$ and the efficiency corrected signal events $N_i$ as free parameters:
\begin{equation}
{\mathscr{L}}=\prod\limits_{k=1}^{N_{\textrm{bins}}} \textrm{Poisson}(n_{\textrm{data},k}, n_{\textrm{exp},k}) \prod\limits_{j=1}^{N_{\textrm{bkg}}}\frac{1}{\sqrt{2\pi}\sigma_{\textrm{bkg,}j}}\exp\left({\frac{-(n_{\textrm{bkg,}j}-\hat n_{\textrm{bkg,}j})^2}{2\sigma ^{2}_{\textrm{bkg,}j}}}\right).
\label{eq:LHFit}
\end{equation}
Here, $n_{\textrm{data},k}$ represents the number of events in each bin $k$. The expected number of background events $\hat n_{\textrm{bkg,}j}$ of each background source $j$ and their normalisation uncertainties $\sigma_{\textrm{bkg,}j}$ are used to constrain the fit. The fit parameters scaling the background contributions are treated as correlated across all channels except for the fake-lepton background, which is uncorrelated across lepton flavours and \bt-tag regions. The size of the background normalisation uncertainties $\sigma_{\textrm{bkg,}j}$ is described in Chapter \ref{sec:uncertainties}.

Combined fits of the \costheta\ distributions using up to four different channels (\ejets\ and \mujets, both with 1 $b$-tag or $\geq$ 2 $b$-tags) are performed for the leptonic and hadronic analyser individually. For each channel, individual templates of the signal and backgrounds are utilised. The combination leading to the lowest total uncertainty is used to quote the result. The helicity fractions are obtained from the fitted values of $n_{i}$ using equations \ref{eq:Frac}--\ref{eq:background}. The fit method is validated using pseudo-experiments varying \FN over the range [0.4, 1.0], \FL over the range [0.15, 0.45] and \FR over the range [-0.15, 0.15]. For each set, the unitarity constraint (\FN + \FL + \FR = 1) is imposed. No bias is observed.

The uncertainties in the helicity fractions obtained from the fit include both the statistical uncertainty of the data and the systematic uncertainty of the background normalisations.
For the leptonic analyser, the most sensitive results are obtained for the two-channel combination (electron~+~muon) in the $\geq$2 $b$-tags region. Adding further channels increases the total systematic uncertainty, in particular due to uncertainties in the $b$-tagging, which do not compensate with the decrease in the statistical uncertainty.
For the hadronic analyser, the four-channel combination (including both the 1 $b$-tag and $\geq$2 $b$-tags regions) improves the sensitivity compared to the two-channel combination. 
For each source of systematic uncertainty, modified pseudo-data templates are created and evaluated via ensemble testing. The differences between the mean helicity fractions measured using the nominal templates and those varied to reflect systematic errors are quoted as systematic uncertainty. Systematic uncertainties from different sources, described in the following section, are treated as uncorrelated.

\section{Systematic uncertainties}
\label{sec:uncertainties}

Systematic uncertainties from several sources can affect the normalisation of the signal and background and/or the shape of the \costheta\ distribution. Correlations of a given systematic uncertainty are maintained across processes and channels, unless otherwise stated. The impact of uncertainties from the various sources is determined using a frequentist method based on the generation of pseudo-experiments.                                                                        

\subsection*{Uncertainties associated with reconstructed objects}
Different sources of systematic uncertainty affect the reconstructed objects used in this analyses. All these sources, described in the following, are propagated to changes in the shape of the \costheta\ distributions.

Uncertainties associated with the lepton selection arise from the trigger, reconstruction, identification and isolation efficiencies, as well as the lepton momentum scale and resolution.
 They are estimated from $Z \rightarrow \ell^{+} \ell^{-} (\ell = e,\,\mu)$, $J/\psi \rightarrow \ell^{+} \ell^{-}$ and $W \rightarrow e\nu$ processes in data and in simulated samples using tag-and-probe techniques described in Refs.~\cite{PERF-2013-03, PERF-2011-01, PERF-2010-04, PERF-2014-05, PERF-2013-05}. 
 Since small differences are observed between data and simulation, correction factors and their related uncertainties are considered to account for these differences. The effect of these uncertainties is propagated through the analysis and represent a minor source of uncertainty in this measurement.

Uncertainties associated with the jet selection arise from the jet energy scale, jet energy resolution, jet vertex fraction requirement and jet reconstruction efficiency. The jet energy scale and its uncertainty are derived combining information from test-beam data, LHC collision data, and simulation~\cite{PERF-2012-01}. The jet energy scale uncertainty is split into 22 uncorrelated sources that have different jet $p_{\textrm{T}}$ and $\eta$ dependencies and are treated independently in this analysis.
 The uncertainty related to the jet energy resolution is estimated by smearing the energy of jets in simulation by the difference between the jet energy resolutions for data and simulation~\cite{PERF-2011-04}.
 The efficiency for each jet to satisfy the jet vertex fraction requirement is measured in $Z \rightarrow \ell^{+} \ell^{-}+\textrm{1-jet}$ events in data and simulation~\cite{PERF-2014-03}. The corresponding uncertainty is evaluated in the analysis by changing the nominal jet vertex fraction cut value and repeating the analysis using the modified cut value \cite{ATLAS-CONF-2013-083}.
 The jet reconstruction efficiency is found to be about $0.2\%$ lower in simulation than in data for jets below $30$~\GeV~and consistent with data for higher jet $p_T$. All jet-related kinematic variables (including the missing transverse momentum) are recomputed by removing randomly $0.2\%$ of the jets with $p_{\textrm{T}}$ below $30$~\GeV~and the event selection is repeated.

Since the $b$-tagging efficiencies and misidentification rates are not modelled satisfactorily in MC simulation, all jets are assigned a specific \pT- and $\eta$-dependent scale factor to account for this difference. The uncertainties in these scale factors are propagated to the measured value.

An additional uncertainty is assigned due to the extrapolation of the $b$-tagging efficiency measurement to the high-\pT region. Twelve uncertainties are considered for the light-jet tagging, all depending on jet \pT and $\eta$. These systematic uncertainties are taken as uncorrelated.

The uncertainties from the energy scale and resolution corrections for leptons and jets are propagated into the $\ET^{\text{miss}}$ calculation.
Additional uncertainties are added to account for contributions from energy deposits not associated with any jet and due to soft-jets ($7~$GeV$~<\pT<~20$~GeV), and are treated as fully correlated with each other. The uncertainty in the description of extra energy deposited due to pile-up interactions is treated as a separate $\ET^{\text{miss}}$ scale uncertainty. This uncertainty has a negligible effect on the measured \Wboson\ boson helicity fractions.

\subsection*{Uncertainties in signal modelling}

The uncertainties in the signal modelling affect the kinematic properties of simulated \ttbar\ events and thus the acceptance and the shape of the reconstructed \costheta\ distribution.

To assess the impact of the different parton shower and hadronisation models, the \powhw\ sample is compared to a \powpy\ sample and the symmetrised difference is taken as a systematic uncertainty. Similarly, an uncertainty due to the matrix element (ME) MC event generator choice for the hard process is estimated by comparing events produced by \POWHEGBOX\ and MC@NLO, both interfaced to \HERWIG\ for showering and hadronisation. 
The uncertainties due to QCD initial- and final-state radiation (ISR/FSR) modelling are estimated using two \powpy\ samples with varied parameters producing more and less radiation. The larger of the changes due to the two variations is taken and symmetrised.

The uncertainty in the \ttbar signal due to the PDF choice is estimated following the
PDF4LHC recommendations~\cite{Botje:2011sn}. It takes into account
the differences between three PDF sets: CT10~NLO,  MSTW2008~68\,\%~CL~NLO and NNPDF~2.3~NLO~\cite{Ball:2012cx}.
The final PDF uncertainty is an envelope of an intra-PDF uncertainty,
which evaluates the changes due to the variation of different PDF parameters
within a single PDF error set, and an inter-PDF uncertainty, which evaluates
differences between different PDF sets. Each PDF set has a prescription to evaluate an overall uncertainty using its error sets: symmetric Hessian in the case of CT10, asymmetric Hessian for MSTW and sample standard deviation in the NNPDF case. Half the width of the envelope of the three estimates is taken as the PDF systematic uncertainty.

The effect of the uncertainty in the top quark mass is estimated using MC samples with different input top masses for the signal process. The dependence of the obtained helicity fractions on the top quark mass is fitted with a linear function. The uncertainties in the helicity fractions are obtained from the slopes multiplied by the uncertainty in the top quark mass of $172.84 \pm 0.70\,\GeV$~\cite{TOPQ-2016-03} measured by ATLAS at $\sqrt{s}$~=~8\,\tev .

\subsection*{Uncertainties in background modelling}
\label{sec:syst_bkgs}

The different flavour samples of the $W$+jets background are scaled by data-driven calibration factors~\cite{TOPQ-2014-16} as explained in Section~\ref{sec:dataAndMCsamples}. 
All sources of uncertainty on the correction factors other than normalisation (e.g. associated with the objects identification, reconstruction and calibration, etc.) are propagated to the $W$+jets estimation. Their normalisation uncertainty (5\,\% for $W$+light-jets, 25\,\% for $W$+$c$-jets and 7\,\% for $W$+$bb$/$cc$) is taken into account in the likelihood fit as explained in Section~\ref{sec:analysis}.

A relative uncertainty of 30\,\%, estimated using various control regions in the matrix method calculation~\cite{ATLAS-CONF-2014-058}, is used for the fake-lepton contribution. 

For single top quark production, a normalisation uncertainty of 17\,\% is assumed, which takes into account the weighted average of the theoretical uncertainties in $s$-, $t$- and $Wt$-channel production (+$5$\,\%/$-4$\,\%) as well as additional uncertainties due to variations in the amount of initial- and final-state radiation and the extrapolation to high jet multiplicity. The uncertainty in the single-top background shape is assessed by comparing $Wt$-channel Monte Carlo samples generated using alternative methods to take into account $Wt$ and \ttbar\ diagrams interference: diagram removal and diagram subtraction~\cite{Sgtop_WtAndTtbar}.

An overall normalisation uncertainty of 48\,\% is applied to $Z$+jets and diboson contributions. It takes into account a 5\,\% uncertainty in the theoretical (N)NLO cross-section as well as the uncertainty associated with the extrapolation to high jet multiplicity (24\,\% per jet).

All normalisation uncertainties are included in the fit of the \Wboson\ boson helicity fractions via priors for the background yields. While the $W$+jets and fake-lepton uncertainties are included directly, the uncertainty in the total remaining background from other sources is combined to 16\,\% (17\,\%) in the $\geq$2 $b$-tags regions (1~$b$-tag~+~$\geq$2 $b$-tags regions) by adding the uncertainties in the theoretical cross-sections of the single top quark, diboson and $Z$+jets contributions in quadrature. The uncertainty in the shape of the $W$+jets background is considered by jet flavour decomposition. Further background shape uncertainties were evaluated and found to be negligible. 

\subsection*{Other uncertainties}
\label{sec:syst_others}
The uncertainty associated with the limited number of MC events in the signal and background templates is evaluated by performing pseudo-experiments on MC events.

The impact of the 1.9\,\% luminosity uncertainty~\cite{Aaboud:2016hhf} 
is found to be negligible since the background normalisations are constrained in the fit.

\section{Results}
\label{sec:results}

The measured \Wbos\ helicity fractions obtained using the leptonic analyser in semileptonic \ttbar\ events with $\geq$2~$b$-tags are presented in Table~\ref{tab:results_lep}.
\begin{table}[h!]
\centering
  \begin{tabular}{l}
\hline
\multicolumn{1}{c}{Leptonic analyser ($\geq$2 $b$-tags) }\\\hline\\[-8pt]
\FN = 0.709 $\pm$ 0.012 \statBG ${}^{+0.015}_{-0.014}$ \syst     \\[5pt]
\FL = 0.299 $\pm$ 0.008 \statBG ${}^{+0.013}_{-0.012}$ \syst    \\[5pt]
\FR = $-0.008$ $\pm$ 0.006 \statBG $\pm 0.012$ \syst    \\[5pt]
\hline
\end{tabular}
\caption{Measured \Wboson~boson helicity fractions obtained from the leptonic analyser including the statistical uncertainty from the fit and the background normalisation as well as the systematic uncertainty.}
\label{tab:results_lep}
\end{table}
By construction, the individual fractions sum up to one.
The \FN value is anti-correlated with both \FL and \FR ($\rho_{F_{\mathrm{0}}, F_{\mathrm{L}}}=-0.82$, $\rho_{F_{\mathrm{0}}, F_{\mathrm{R}}}=-0.97)$, and \FL and \FR are positively correlated ($\rho_{F_{\mathrm{L}}, F_{\mathrm{R}}}=+0.65)$. The quoted values correspond to the total correlation coefficient, considering statistical and systematic uncertainties.
These results are the most precise \Wbos\ helicity fractions measured so far and are consistent with the SM predictions given at NNLO accuracy~\cite{WHel_theo}. The inclusion of single~$b$-tag regions does not improve the sensitivity, due to larger systematic uncertainties.

The \Wbos\ helicity fractions obtained using the hadronic analyser of semileptonic \ttbar\ events with 1~$b$-tag and $\geq$~2$b$-tags are given in Table~\ref{tab:results_had}. Using the hadronic analyser, the correlations between the helicity fraction are $\rho_{F_{\mathrm{0}}, F_{\mathrm{L}}}$~=~0.56, $\rho_{F_{\mathrm{0}}, F_{\mathrm{R}}}$~=~-0.91 and $\rho_{F_{\mathrm{L}}, F_{\mathrm{R}}}$~=~-0.92.The large anticorrelation between \FL and \FR is a consequence of the low separation power between the up- and down-type quark from the $W$ decay and the resulting similar shapes of the templates of \FL and \FR (see Figure \ref{fig:templates}). The results obtained with the two analysers agree well. The combination of leptonic and hadronic analysers has been tested and, despite the improvement in the statistical uncertainty, it does not improve the total uncertainty.
\begin{table}[h!]
  \centering
  \begin{tabular}{l}
\hline
\multicolumn{1}{c}{Hadronic analyser (1 $b$-tag~+~$\geq$2 $b$-tags)}\\\hline\\[-8pt]
\FN = 0.659 $\pm$ 0.010 \statBG ${}^{+0.052}_{-0.054}$ \syst     \\[5pt]
\FL = 0.281 $\pm$ 0.021 \statBG ${}^{+0.063}_{-0.067}$ \syst    \\[5pt]
\FR = 0.061 $\pm$ 0.022 \statBG ${}^{+0.101}_{-0.108}$ \syst   \\[5pt]
\hline
\end{tabular}
\caption{Measured \Wboson~boson helicity fractions for the hadronic analyser including the statistical uncertainty from the fit and the background normalisation as well as the systematic uncertainty.}
\label{tab:results_had}
\end{table}

Figure~\ref{fig:postfit_lep} shows, separately for the $e$+jets and $\mu$+jets channels, the distributions of \costheta\ from the leptonic analyser. The distributions for the hadronic analyser are presented in Figure~\ref{fig:postfit_had}. The uncertainty band in the data-to-best-fit ratio represents the statistical and background normalisation uncertainty. The deviations observed in the ratio are covered by the systematic uncertainties. The peak at $\costheta \approx -0.7$ as seen in the single $b$-tag channels in Figure~\ref{fig:postfit_had} is caused by misreconstructed events. A missing second $b$-tag increases the probability of swapping the $b$-quark jet from the top quark decay with the up-type quark jet from the $W$ decay.
\begin{figure}[!ht]
\begin{center}
\includegraphics[width=0.8\textwidth]{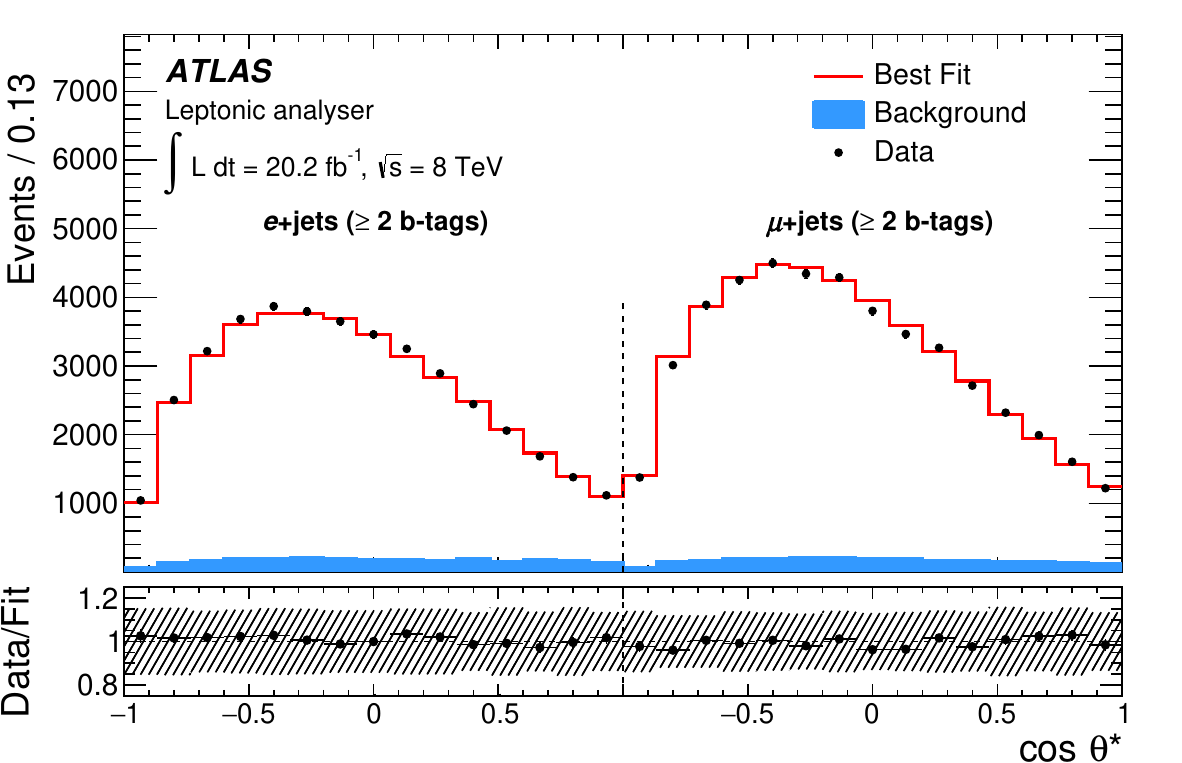}
\caption{Post-fit distribution of \costheta\ for the leptonic analyser with $\geq$2 $b$-tags, in which a two-channel combination is performed (electron and muon). The uncertainty band represents the total uncertainty in the fit result.}
\label{fig:postfit_lep}
\end{center}
\end{figure}
\begin{figure}[!ht]
\begin{center}
\includegraphics[width=1\textwidth]{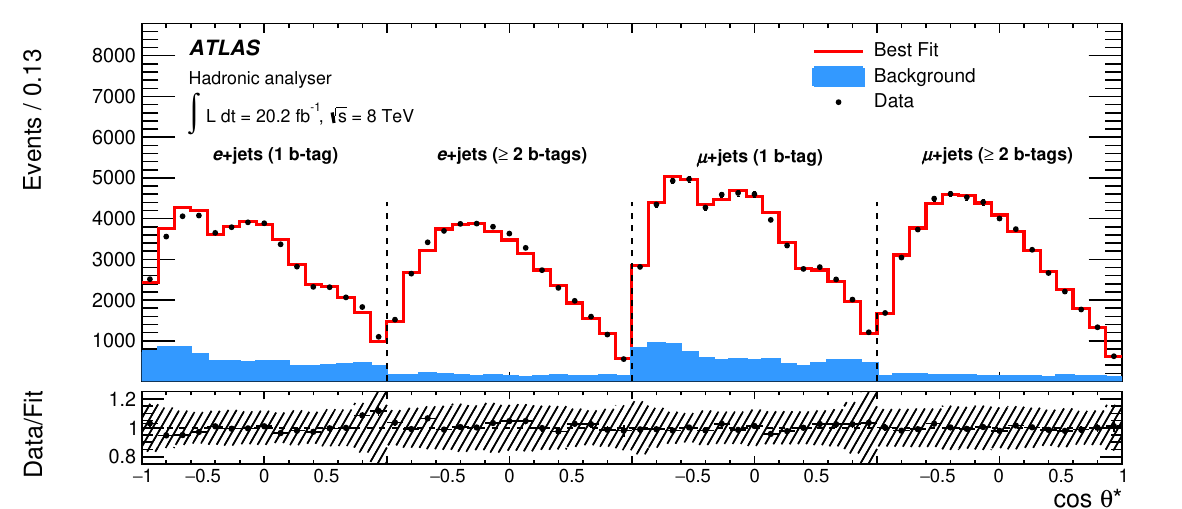}
\caption{Post-fit distribution of \costheta\ for the hadronic analyser, in which the combination of four channels is performed (electron and muon, with exactly 1~$b$-tag and $\geq$2~$b$-tags). The uncertainty band represents the total uncertainty in the fit result.}
\label{fig:postfit_had}
\end{center}
\end{figure}

The contributions of the various systematic uncertainties are quoted in Table~\ref{tab:uncertainties}.
In the case of the leptonic analyser, the dominant contributions come from the jet energy scale and resolution and the statistical error in the MC templates.
For the hadronic analyser, the systematic uncertainties are larger. Including the 1~$b$-tag region aids in reducing the error. One of the main contributions is the \bt-tagging uncertainty, affecting both the event selection and $b$-tag categorisation, as well as the up- vs down-type quark separation. Other major contributions come from the jet energy resolution and the modelling of \ttbar\ events (initial- and final-state radiation, parton showering and hadronisation, and Monte Carlo event generator choice for the matrix elements).
\sisetup{retain-explicit-plus=true, group-digits = false, table-figures-decimal = 4, table-format=1.4, round-minimum = 0.0001, table-space-text-pre = $\pm$,}
\begin{table}[h!]
\centering
\begin{tabular}{
l
S
S
S|@{\quad}
S
S
S
}
\hline\hline
Uncertainty & \multicolumn{3}{c}{Leptonic, $\geq$2 $b$-tags} & \multicolumn{3}{c}{Hadronic, $1\,+\geq$2 $b$-tags} \\\hline\hline
 & \multicolumn{1}{c}\FN & \multicolumn{1}{c}\FL & \multicolumn{1}{c}\FR &  \multicolumn{1}{c}\FN & \multicolumn{1}{c}\FL & \multicolumn{1}{c}\FR \\\hline\hline
{}& \multicolumn{6}{c}{Reconstructed objects} \\\hline\hline

\multirow{2}{*}{Electron} & +0.0028 & +0.0018 & +0.0011 & +0.0025 & +0.0028 & +0.0051\\
                          & -0.0030 & -0.0020 & -0.0011 & -0.0021 & -0.0038 & -0.0058\\\hline
\multirow{2}{*}{Muon} & +0.0024 & +0.0013 & +0.0010 & +0.0026 & +0.0046 & +0.0072\\
                          & -0.0029 & -0.0015 & -0.0015 & -0.0037 & -0.0035 & -0.0072\\\hline
\multirow{2}{*}{Jet energy scale} & +0.0063 & +0.0028 & +0.0037 & +0.0069 & +0.012 & +0.014\\
                          & -0.0033 & -0.0025 & -0.0014 & -0.0070 & -0.008 & -0.005\\\hline
\multirow{2}{*}{Jet energy resolution} & +0.0062 & +0.0048 & +0.0072 & +0.027 & +0.033 & +0.057\\
                          & -0.0059 & -0.0018 & -0.0067 & -0.031 & -0.041 & -0.071\\\hline
\multirow{2}{*}{Jet vertex fraction} & +0.0036 & +0.0019 & +0.0017 & +0.013 & +0.0012 & +0.011\\
                          & -0.0017 & -0.0013 & -0.0006 & -0.009 & -0.0046 & -0.005\\\hline
\multirow{2}{*}{Jet reconstruction efficiency} & +0.0002 & {<0.0001} & +0.0002 & +0.0008 & +0.0004 & +0.0011\\
                          & -0.0002 & {<0.0001} & -0.0002 & -0.0008 & -0.0004 & -0.0011\\\hline
\multirow{2}{*}{\bt-tagging} & +0.0017 & +0.0012 & +0.0011 & +0.029 & +0.013 & +0.034\\
                          & -0.0021 & -0.0013 & -0.0012 & -0.031 & -0.014 & -0.035\\
\hline\hline
\multirow{2}{*}{Sum reconstructed objects} & +0.010 & +0.0064 & +0.0085 & +0.043 & +0.038 & +0.069\\
                          & -0.008 & -0.0044 & -0.0072 & -0.045 & -0.044 & -0.080\\\hline
\hline
{}& \multicolumn{6}{c}{Signal modelling} \\\hline\hline
{Showering and hadronisation} & {$\pm$}0.0019 & {$\pm$}0.0019 & {$\pm$}0.0037& {$\pm$}0.015& {$\pm$}0.001 & {$\pm$}0.014\\\hline
{ME event generator} & {$\pm$}0.0025 & {$\pm$}0.0032 & {$\pm$}0.0057 & {$\pm$}0.016 & {$\pm$}0.024 & {$\pm$}0.040\\\hline
{ISR/FSR} & {$\pm$}0.0033 & {$\pm$}0.0058 & {$\pm$}0.0034 & {$\pm$}0.018 & {$\pm$}0.039 & {$\pm$}0.057\\\hline
{PDF} & {$\pm$}0.0033 & {$\pm$}0.0042 & {$\pm$}0.0009 & {$\pm$}0.0010 & {$\pm$}0.0020 & {$\pm$}0.0020\\\hline
{Top quark mass} & {$\pm$}0.0017 & {$\pm$}0.0050 & {$\pm$}0.0033 & {$\pm$}0.0033 & {$\pm$}0.0100 & {$\pm$}0.0068\\
\hline\hline
{Sum signal modelling} & {$\pm$}0.0058 & {$\pm$}0.0094 & {$\pm$}0.0082 & {$\pm$}0.028 & {$\pm$}0.047 & {$\pm$}0.072\\
\hline\hline
{}& \multicolumn{6}{c}{Method uncertainty} \\\hline\hline
{Template statistics} & {$\pm$}0.0091 & {$\pm$}0.0056 & {$\pm$}0.0044 & {$\pm$}0.0076 & {$\pm$}0.016 & {$\pm$}0.016 \\
\hline\hline
{}& \multicolumn{6}{c}{Total uncertainty} \\\hline\hline
\multirow{2}{*}{Total systematic} & +0.015 & +0.013 & +0.013 & +0.052 & +0.063 & +0.100\\
                          & -0.014 & -0.012 & -0.012 & -0.054 & -0.067 & -0.110\\\hline\hline
Stat. + bkg. norm& {$\pm$}0.012 & {$\pm$}0.008 & {$\pm$}0.006 & {$\pm$}0.010 & {$\pm$}0.021 & {$\pm$}0.022\\

\hline\hline
\end{tabular}
\caption{Summary of systematic and statistical uncertainties for the measurements obtained using the leptonic (left) and the hadronic (right) analysers. The numbers in the last row (Stat. + bkg. norm) correspond to the statistical uncertainty of the fit, including the normalisation uncertainties in the background yields.}
\label{tab:uncertainties}
\end{table}

\begin{table}[!h]
  \centering
  \begin{tabular}{c | l}
Coupling & 95\,\% CL interval\\
\hline
\hline
  $V_{\text{R}}$ & $[-0.17, 0.25]$\\
  $g_{\text{L}}$ & $[-0.11, 0.08]$\\
   $g_{\text{R}}$ & $[-0.03, 0.06], [0.74, 0.78]$\\
\end{tabular}
\caption{Allowed ranges for the anomalous couplings $V_{\text{R}}$, $g_{\text{L}}$, and $g_{\text{R}}$ at 95\,\%~CL. The limits are derived using the measured \Wboson\ helicity fractions using the leptonic analyser for events with $\geq$2~$b$-tags (combination of the two channels, electron and muon).}
\label{tab:1d_limits}
\end{table}

Within the effective field theory framework~\cite{AguilarSaavedra:2006fy}, the $Wtb$ decay vertex can be parameterised in terms of anomalous couplings as shown in Equation~\eqref{eq:Wtb}. Limits on these anomalous left- and right-handed vector and tensor couplings are set using the EFTfitter tool~\cite{EFTFitter} and the model of \cite{AguilarSaavedra:2006fy}. The anomalous couplings are assumed to be real, corresponding to the CP-conserving case. As the \Wboson\ helicity fractions only allow the ratios of couplings to be constrained, the value of $V_{\mathrm{L}}$  is fixed to the Standard Model prediction of one. The correlations of systematic uncertainties are taken into account. Figure~\ref{fig:limits} shows the limits on $g_{\mathrm{L}}$ and $g_{\mathrm{R}}$ couplings while $V_{\mathrm{L}}$ and $V_{\mathrm{R}}$ are fixed to their SM values, as well as $V_{\mathrm{R}}$ and $g_{\mathrm{R}}$ limits, where the other couplings are fixed to their SM values. The intervals are obtained using the leptonic analyser since it provides the most sensitive results. Table \ref{tab:1d_limits} shows the 95\,\% confidence level (CL) intervals for each anomalous coupling while fixing all others to their SM value. These limits correspond to the set of smallest intervals containing 95\,\% of the marginalised posterior distribution for the corresponding parameter.
\begin{figure}[!h]
  \begin{center}
              \mbox{
    \subfloat[]{\includegraphics[height=70mm]{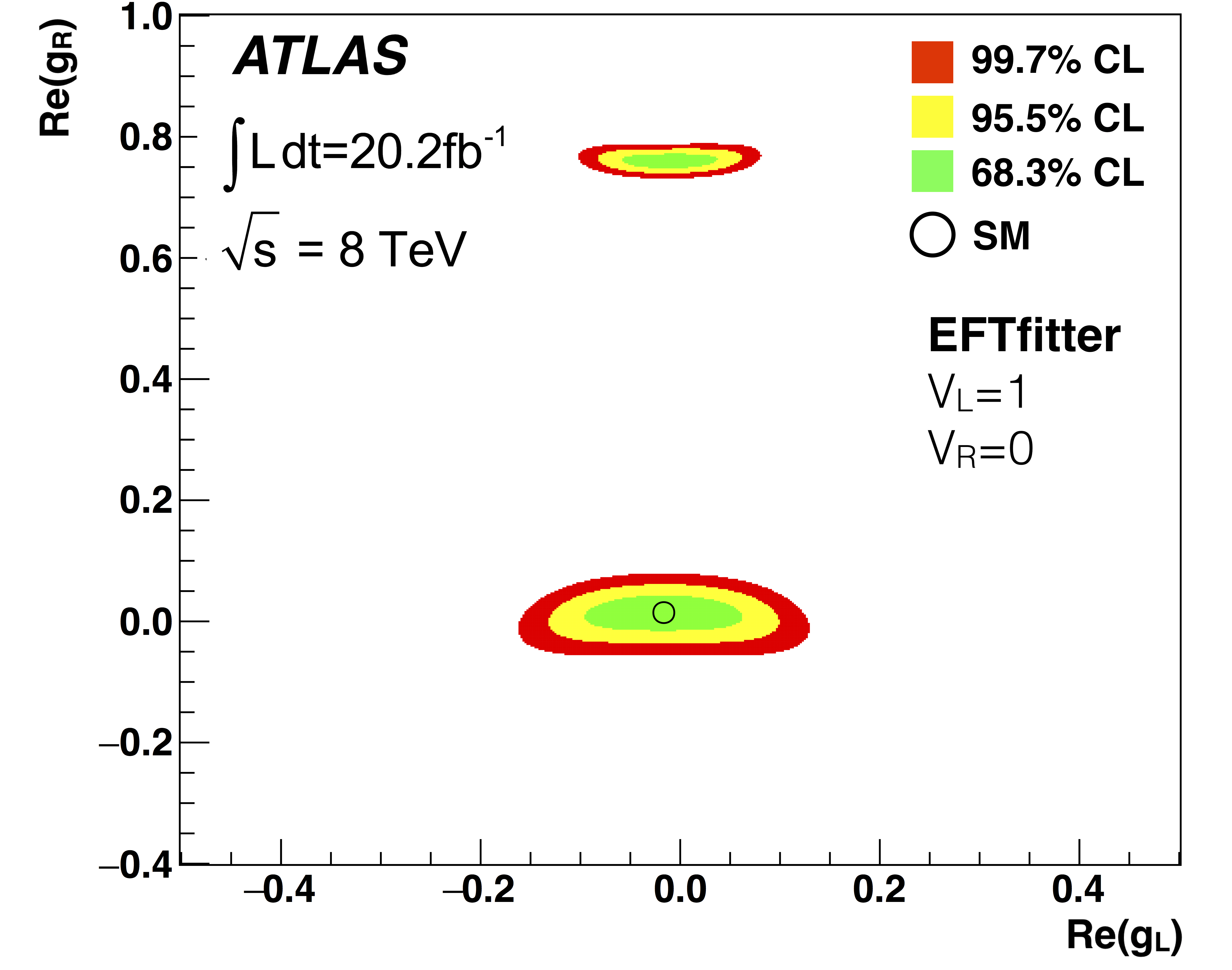}\label{fig:gl_gr_limits}}\quad
    \subfloat[]{\includegraphics[height=70mm]{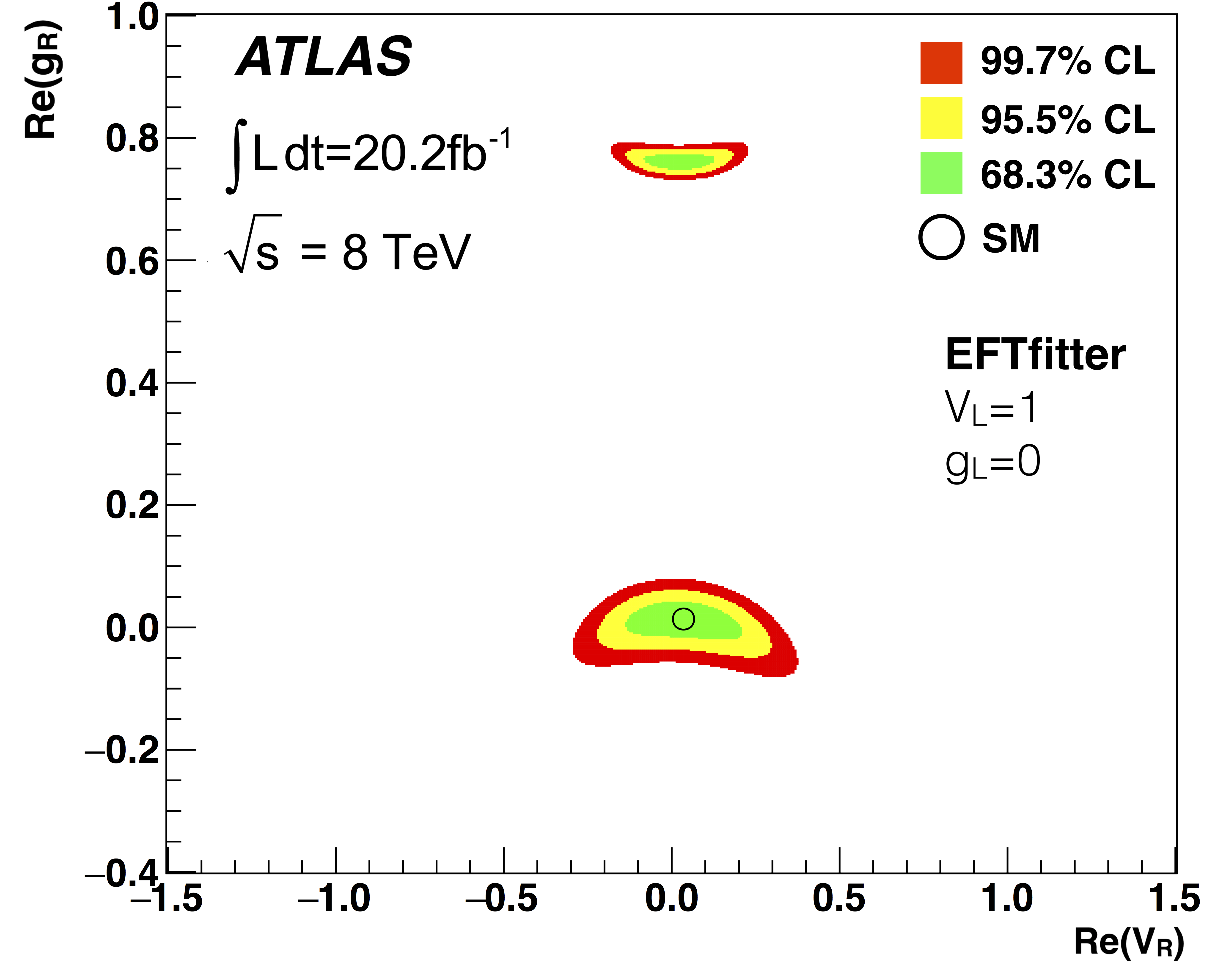}\label{fig:VR_gr_limits}}\quad
  }
    \caption{\protect\subref{fig:gl_gr_limits} Limits on the anomalous left- and right-handed tensor couplings of the $Wtb$ decay vertex as obtained from the measured \Wbos\ helicity fractions from the leptonic analyser. \protect\subref{fig:VR_gr_limits} Limits on the right-handed vector and tensor coupling. As the couplings are assumed to be real, the real part corresponds to the magnitude. Unconsidered couplings are fixed to their SM values.}
    \label{fig:limits}
  \end{center}
\end{figure}
Similar limits on the anomalous couplings were derived by both the ATLAS and CMS experiments using the measured helicity fractions of \Wboson\ bosons~\cite{TOPQ-2011-10, CMS-TOP-13-008}. Complementary limits can be set by other measurements: the allowed region of $g_{\mathrm{R}} \approx 0.75$ is excluded by measurements of the $t$-channel single top quark production~\cite{EFTFitter, TOPQ-2013-12, CMS-TOP-12-038, TOPQ-2012-21} which also constrains $V_{\mathrm{L}}$. The branching fraction of $\bar{B} \rightarrow X_s \gamma$ allow more stringent limits to be set on $g_{\mathrm{L}}$ and $V_{\mathrm{R}}$~\cite{Grzadkowski:2008mf}.

\section{Conclusion}
\label{sec:conclusion}

The longitudinal, left- and right-handed \Wboson~boson helicity fractions are measured using the angle between the charged lepton (\dQ) and the reversed \bquark\ direction in the \Wboson~boson rest frame for leptonically (hadronically) decaying \Wboson~bosons from \ttbar\ decays. A data set corresponding to 20.2 \ifb\ of $pp$ collisions at the LHC with a \com\ energy of \rts = 8 \tev, recorded by the ATLAS experiment, is analysed. Events are required to include one isolated electron or muon and at least four jets, with at least one of them tagged as a \bt-jet. Events are reconstructed using a kinematic likelihood fit based on mass constraints for the top quarks and \Wboson\ bosons. It utilises the weight of the \bt-jet tagging algorithm to further separate the up- and down-type quarks from the hadronically decaying \Wboson\ bosons. The fractions for left-handed, right-handed and longitudinally polarised \Wboson\ bosons are found to be \FN = \focent\ $\pm$ \fostat\ \statBG ${}\pm{\fosystplus}$ \syst, \FL = \flcent\ $\pm$ \flstat\ \statBG ${}\pm{\flsystplus}$ \syst and \FR = \frcent\ $\pm$ \frstat\ \statBG ${}\pm{\frsystplus}$ \syst. 
These results constitute the most precise measurement of the \Wboson\ helicity fractions in \ttbar\ events to date and are in good agreement with the Standard Model predictions within uncertainties. Using these results, limits on anomalous couplings of the $Wtb$ vertex are set.

\clearpage

\section*{Acknowledgements}

We thank CERN for the very successful operation of the LHC, as well as the
support staff from our institutions without whom ATLAS could not be
operated efficiently.

We acknowledge the support of ANPCyT, Argentina; YerPhI, Armenia; ARC, Australia; BMWFW and FWF, Austria; ANAS, Azerbaijan; SSTC, Belarus; CNPq and FAPESP, Brazil; NSERC, NRC and CFI, Canada; CERN; CONICYT, Chile; CAS, MOST and NSFC, China; COLCIENCIAS, Colombia; MSMT CR, MPO CR and VSC CR, Czech Republic; DNRF and DNSRC, Denmark; IN2P3-CNRS, CEA-DSM/IRFU, France; SRNSF, Georgia; BMBF, HGF, and MPG, Germany; GSRT, Greece; RGC, Hong Kong SAR, China; ISF, I-CORE and Benoziyo Center, Israel; INFN, Italy; MEXT and JSPS, Japan; CNRST, Morocco; NWO, Netherlands; RCN, Norway; MNiSW and NCN, Poland; FCT, Portugal; MNE/IFA, Romania; MES of Russia and NRC KI, Russian Federation; JINR; MESTD, Serbia; MSSR, Slovakia; ARRS and MIZ\v{S}, Slovenia; DST/NRF, South Africa; MINECO, Spain; SRC and Wallenberg Foundation, Sweden; SERI, SNSF and Cantons of Bern and Geneva, Switzerland; MOST, Taiwan; TAEK, Turkey; STFC, United Kingdom; DOE and NSF, United States of America. In addition, individual groups and members have received support from BCKDF, the Canada Council, CANARIE, CRC, Compute Canada, FQRNT, and the Ontario Innovation Trust, Canada; EPLANET, ERC, ERDF, FP7, Horizon 2020 and Marie Sk{\l}odowska-Curie Actions, European Union; Investissements d'Avenir Labex and Idex, ANR, R{\'e}gion Auvergne and Fondation Partager le Savoir, France; DFG and AvH Foundation, Germany; Herakleitos, Thales and Aristeia programmes co-financed by EU-ESF and the Greek NSRF; BSF, GIF and Minerva, Israel; BRF, Norway; CERCA Programme Generalitat de Catalunya, Generalitat Valenciana, Spain; the Royal Society and Leverhulme Trust, United Kingdom.

The crucial computing support from all WLCG partners is acknowledged gratefully, in particular from CERN, the ATLAS Tier-1 facilities at TRIUMF (Canada), NDGF (Denmark, Norway, Sweden), CC-IN2P3 (France), KIT/GridKA (Germany), INFN-CNAF (Italy), NL-T1 (Netherlands), PIC (Spain), ASGC (Taiwan), RAL (UK) and BNL (USA), the Tier-2 facilities worldwide and large non-WLCG resource providers. Major contributors of computing resources are listed in Ref.~\cite{ATL-GEN-PUB-2016-002}.

\clearpage

\printbibliography
\newpage
\begin{flushleft}
{\Large The ATLAS Collaboration}

\bigskip

M.~Aaboud$^\textrm{\scriptsize 137d}$,
G.~Aad$^\textrm{\scriptsize 88}$,
B.~Abbott$^\textrm{\scriptsize 115}$,
J.~Abdallah$^\textrm{\scriptsize 8}$,
O.~Abdinov$^\textrm{\scriptsize 12}$,
B.~Abeloos$^\textrm{\scriptsize 119}$,
O.S.~AbouZeid$^\textrm{\scriptsize 139}$,
N.L.~Abraham$^\textrm{\scriptsize 151}$,
H.~Abramowicz$^\textrm{\scriptsize 155}$,
H.~Abreu$^\textrm{\scriptsize 154}$
% ...
\bigskip
\\
% ...
$^{8}$ Department of Physics, The University of Texas at Arlington, Arlington TX, United States of America\\
% ...
$^{12}$ Institute of Physics, Azerbaijan Academy of Sciences, Baku, Azerbaijan\\
% ...
$^{88}$ CPPM, Aix-Marseille Universit{\'e} and CNRS/IN2P3, Marseille, France\\
% ...
$^{115}$ Homer L. Dodge Department of Physics and Astronomy, University of Oklahoma, Norman OK, United States of America\\
% ...
$^{119}$ LAL, Univ. Paris-Sud, CNRS/IN2P3, Universit{\'e} Paris-Saclay, Orsay, France\\
% ...
$^{137}$ $^{(a)}$ Facult{\'e} des Sciences Ain Chock, R{\'e}seau Universitaire de Physique des Hautes Energies - Universit{\'e} Hassan II, Casablanca; $^{(b)}$ Centre National de l'Energie des Sciences Techniques Nucleaires, Rabat; $^{(c)}$ Facult{\'e} des Sciences Semlalia, Universit{\'e} Cadi Ayyad, LPHEA-Marrakech; $^{(d)}$ Facult{\'e} des Sciences, Universit{\'e} Mohamed Premier and LPTPM, Oujda; $^{(e)}$ Facult{\'e} des sciences, Universit{\'e} Mohammed V, Rabat, Morocco\\
% ...
$^{139}$ Santa Cruz Institute for Particle Physics, University of California Santa Cruz, Santa Cruz CA, United States of America\\
% ...
$^{151}$ Department of Physics and Astronomy, University of Sussex, Brighton, United Kingdom\\
% ...
$^{154}$ Department of Physics, Technion: Israel Institute of Technology, Haifa, Israel\\
$^{155}$ Raymond and Beverly Sackler School of Physics and Astronomy, Tel Aviv University, Tel Aviv, Israel\\
% ...
$^{a}$ Also at Department of Physics, King's College London, London, United Kingdom\\
$^{b}$ Also at Institute of Physics, Azerbaijan Academy of Sciences, Baku, Azerbaijan\\
$^{c}$ Also at Novosibirsk State University, Novosibirsk, Russia\\
$^{d}$ Also at TRIUMF, Vancouver BC, Canada\\
$^{e}$ Also at Department of Physics {\&} Astronomy, University of Louisville, Louisville, KY, United States of America\\
% ...
\end{flushleft}


\end{document}